\documentclass[11pt]{amsart}

\usepackage{setspace}
\usepackage{graphicx}
\usepackage{amssymb,amsthm,amsmath}
\usepackage[font=onehalfspacing]{caption}
\usepackage{subcaption}
\usepackage{xcolor}
\usepackage{tabu}
\usepackage{physics}
\usepackage{mathtools}
\usepackage{url}
\usepackage{multirow}
\usepackage{hyperref}
\usepackage{cite}
\usepackage{enumitem}
\usepackage{amsfonts}
\usepackage{lineno}
\usepackage[utf8]{inputenc} 
\usepackage[autostyle]{csquotes}
\MakeOuterQuote{"}

\DeclarePairedDelimiter\floor{\lfloor}{\rfloor}
\newtheorem{theorem}{Theorem}

\calclayout

\newcommand{\modi}[1]{\textcolor{black}{#1}}

\begin{document}

\title{Extracting Complements and Substitutes from Sales Data: A Network Perspective}
\xdef\shorttitle{Extracting Complements and Substitutes: A Network Perspective}
\xdef\shortauthors{Tian et al.}
\maketitle

\begin{center}
	Yu Tian\textsuperscript{1}, Sebastian Lautz\textsuperscript{2}, Alisdair \hspace{-.1em}O. \hspace{-.2em}G. \hspace{-.1em}Wallis\textsuperscript{2} and Renaud Lambiotte\textsuperscript{1} \par \bigskip
	
	\textsuperscript{1}\textit{\small{Mathematical Institute, University of Oxford, Oxford, OX2 6GG}} \par
	\textsuperscript{2}\textit{\small{Tesco PLC, Tesco House, Shire Park, Welwyn Garden City, AL7 1GA}}
\end{center}

\begin{abstract} 
The complementarity and substitutability between products are essential concepts in retail and marketing. Qualitatively, two products are said to be substitutable if a customer can replace one product by the other, while they are complementary if they tend to be bought together.
In this article, we take a network perspective to help automatically identify complements and substitutes from sales transaction data. 
Starting from a bipartite product-purchase network representation, with both transaction nodes and product nodes, we develop appropriate null models to infer significant relations, either complements or substitutes, between products, and design measures based on random walks to quantify their importance. The resulting unipartite networks between products are then analysed with community detection methods, in order to find groups of similar products for the different types of relationships. The results are validated by combining observations from a real-world basket dataset with the existing product hierarchy, as well as a large-scale flavour compound and recipe dataset. 
\end{abstract}


\section{Introduction}
Understanding the hidden relations existing between products is fundamental in both economics and marketing research as well as in retail \cite{elrod_inferring_2002}. This question lies at the core of market structure analysis and finds numerous applications. Retailers must regularly make decisions taking product relationships into account \cite{manrali_assortment_2009}, for instance to design their product catalogue and to determine the number of products to offer in each category \cite{Kok_assortplanning_2015}. Brick-and-mortar retailers seek to identify the best way to arrange the product layout in aisles and stock their shelves \cite{nierop_shelf_2008}, and online retailers also strive to optimise the grouping of products in their online shops \cite{Breugelmans_online-retail_2007}. Furthermore, they must \modi{decide} which products to bundle or promote together. These assortment-related decisions have significant influence on customers' choices, sales of products, and finally, profits \cite{manrali_assortment_2009, Kok_assortplanning_2015, briesch_choice_2009}. 

\textit{Complements} and \textit{substitutes} are two central concepts to characterise relationships between products, with well-established definitions in economics \cite{nilcholson_microeco_2012}. Complementary products are sold separately but used together, each creating a demand for the other, such as hot dogs and hot dog buns. Substitute products serve the same purpose and can be used in place of one another, such as Brand A tomatoes and Brand B tomatoes. \modi{In the economics literature}, the degree of complementarity (substitutability) is formally defined \modi{through} the negative (positive) cross-price elasticity of demand, where the rise in the demand of one product is recorded after the price of the other product is reduced (increased) by a unit. The mechanisms of complements and substitutes are also referred to as the \textit{halo effect} and \textit{demand transfer}, respectively, in the retail context \cite{ailawadi_halo_2007}.

Despite its practical importance, the algorithmic problem of identifying the relationship between products in retail is not well known. For a long time, researchers and practitioners have selected the set of possible complementary or substitute products by means of, for instance, field expertise and simple statistics, and the analysis has usually been restricted to a fairly small number of products \cite{song_discretecontinuous_2007, berry_structural_2014}. Recent development and application of natural language processing and machine learning (ML) algorithms, especially those based on word embedding, bring in new visions and opportunities, which makes it possible to analyse thousands of products \cite{gabel_p2vmap_2019, ruiz_shopper_2019, chen_studying_2020}. These methods use the transactions, some require customer information, as the original feature space, and apply ML algorithms to essentially reduce the dimension of these feature vectors. The resulting reduced dimensional embeddings can then be used to identify the relationship between products and also in customer choice models. 

However, there are several limitations in these applications. Firstly, the interpretation of the selected features in the related ML algorithms is difficult. This makes it challenging to develop metrics in this space, in particular to verify the property of triangle inequality, and further use metric invariants to define measures between products. In practice, these methods often rely on the definition of similarity measures (not necessarily metrics) for complementarity and exchangeability\footnote{\footnotesize{Exchangeability is an additional concept in order to verify whether two products are substitutes, and \modi{researchers} postulate that they are if they have low complementarity and high exchangeability.}} \cite{ruiz_shopper_2019, chen_studying_2020}. Secondly, these methods lack specific criteria to determine whether two products are complements, substitutes, or just independent, despite trying to quantify the effects by their similarities. Thirdly, they do not explore the \modi{connection} between complements and substitutes, which is of great significance to improve the understanding of the relationships and their further applications. Furthermore, these methods are based on co-purchase patterns, but the other valuable information in the sales data remains unused. 

In our study, we \modi{propose an alternative to the classic approach based on cross-price elasticity, and take instead} a network perspective in order to define complementarity and substitutability between products. As a starting point, we model the sales data as a bipartite \textit{product-purchase network}, with both transactions (or baskets) and products as nodes. We then perform our analysis directly on the network, without having to rely on low-dimensional embeddings which may lead to uncontrolled loss of information. We use the connectivity patterns between products to characterise complements and substitutes. To do so, we define null models on the bipartite network to determine significant relationships between products, and propose measures induced by random walks on networks to quantify the intensity of these relationships. This approach can be seen as a generalisation of the classic \textit{bipartite network projection} where we focus on different notions of connectivity induced by the bipartite structure. We also take an initial step to explicitly incorporate noise effects in our measures. As we show later, the resulting projections onto unipartite networks, based both on complementary and substitute connections, allow us to find groups of similar products with standard tools like community detection.  

The aim of our work is to provide insights into product relationships from a network perspective with simple assumptions, and to further extract both complements and substitutes efficiently from easily accessible sales data. It is our belief that a network approach opens up a promising new angle on this problem due to its flexibility, e.g. in determining significant relationships, and the vast network science toolbox. The insights derived from our methods have applications in assortment-related decision making, not only for retailers but also general firms with long product lines. Furthermore, our set of methods can also be applied to other contexts, such as trading networks, ecological systems and social networks, where both \modi{the identification of cooperative and competitive relations} are of interest.

\section{Data}
\modi{In this section, we present the different datasets that we use to perform our analysis, the sales data in Sect. \hspace{-.2em}\ref{sec:sales_data} to extract the product relationships, the product hierarchy data in Sect. \hspace{-.2em}\ref{sec:hierarchy_data} and the flavour compound and recipe data in Sect. \hspace{-.2em}\ref{sec:flavour_recipe_data} to validate the results.}

\subsection{Sales data}
\label{sec:sales_data}
We used anonymised grocery sales data from Tesco, the UK's largest supermarket chain. The data consists of timestamped transactions of stores, and it has been anonymised for general research purposes, i.e. each customer's personal identifiable information has been removed. For each store, the transaction data comprises a transaction ID, which gives a unique code to each shopping trip, the date when the transaction was made, the product IDs, and their purchased quantities; \modi{see the top of Fig. \hspace{-.2em}\ref{fig:schem_data_struct}}. 

The data used for this study is from a generic convenience store in \modi{an} urban area, and spans a three-month period avoiding major holidays such as Christmas and Easter. The time window is chosen to be long enough to be representative of the underlying customer population's product purchase patterns, but also sufficiently short to avoid seasonal effects as well as change of behaviour over time. Furthermore, to facilitate the interpretation of the results, we restrict our analysis to fresh fruit, vegetables and salads where we believe complementary and substitute products commonly exist. We also exclude products that are purchased less than once a month, and those in almost every transaction. These result in the final dataset of $43837$ transactions and $253$ products.  

\subsection{Product hierarchy data}
\label{sec:hierarchy_data}
In retail, it is common to organise products in a hierarchy, where similar products are grouped into increasingly generic categories. Products that are close together in the hierarchy are typically sold next to each other in a store. At the lowest hierarchical level, each unique code corresponds to a different product, including the same products of different sizes or flavours. Overall, we have $4$ levels, from \texttt{L1} to \texttt{L4} (excluding the product level). The higher a level is, the more generic the corresponding category. For example, "apple" is a category in the \texttt{L1} hierarchy, and "fruit" is a category in the \texttt{L3} hierarchy. Hence, a natural way to validate our product relationships and to explore their features would be to compare them to the corresponding product hierarchy. 

\subsection{Flavour compound and recipe data}
\label{sec:flavour_recipe_data}
\begin{figure}
    \centering
    \includegraphics[height=.9\textheight]{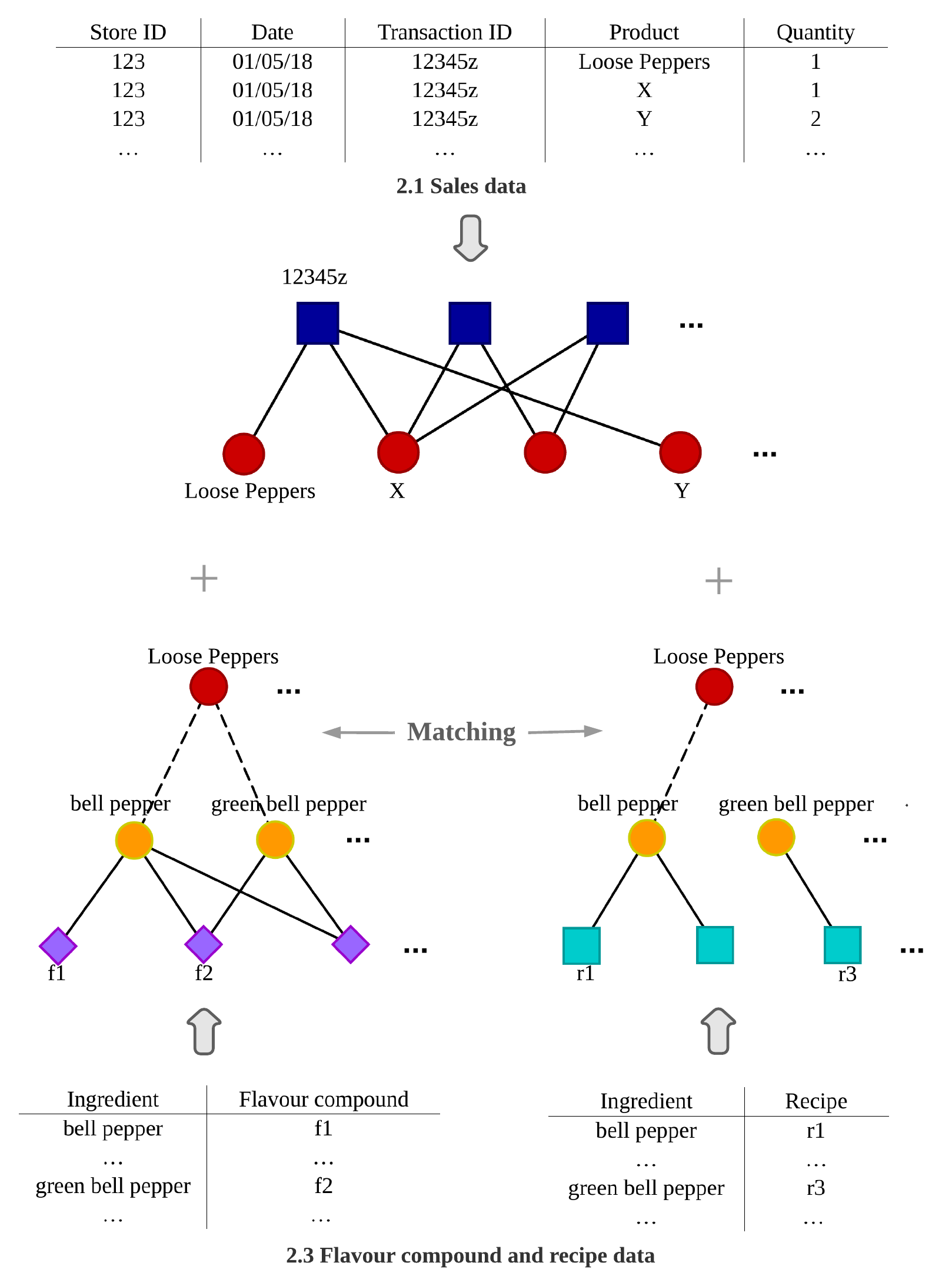}
    \caption{\modi{Schematic diagram showing the data structure of the sales data in Sect. \hspace{-.2em}\ref{sec:sales_data} and the flavour compound and recipe data in Sect. \hspace{-.2em}\ref{sec:flavour_recipe_data}, and their corresponding bipartite networks, together with the matching between the products and the ingredients. The "Matching" step is required because of the different names that can appear in different datasets. }}
    \label{fig:schem_data_struct}
\end{figure}

Ahn et al \cite{Ahn_flavor_2011} provide a systematic list of $1107$ flavour compounds and their natural occurrences in terms of $1525$ ingredients overall from Fenaroli's handbook of flavour ingredients \cite{Fenaroli_flavor_2004}. They also provide $56498$ recipes belonging to geographically distinct cuisines (North American, Western European, Southern European, Latin American and East Asian), which were obtained from \textit{epicurious.com}, \textit{allrecipes.com} and \textit{menupan.com}; see the bottom of Fig. \hspace{-.2em}\ref{fig:schem_data_struct}. Hence, to validate our results from the features in both flavour compounds and recipes, we match our products to their ingredients. 

To construct the correspondence between our products and the flavour compounds, we match each product to as many ingredients as possible. For example, "Loose Peppers" is matched to all possibly equivalent peppers including "bell pepper" and "green bell pepper"; \modi{see the middle left of Fig. \hspace{-.4em}\ref{fig:schem_data_struct}}. This results in our ingredients of interest to be $140$, with their corresponding flavour compounds being $865$, and each ingredient is linked to $57$ flavour compounds on average. Note that there are $11$ products which do not have exactly matched ingredients, hence we match them to generic ones\footnote{\footnotesize{In the recipe data, there are both specific ingredients, e.g. "apple", and generic ones, e.g. "fruit".}}. For example, we match the product "Single Pomegranate" to the ingredient "fruit". There are also $44$ complex products whose ingredients cannot be directly inferred from their names, thus we match them to their main ingredients on the website. For example, we match the product "Cheddar Coleslaw" to the ingredients "cheddar cheese", "cabbage", "carrot" and "onion". 

For the recipe data, we match each product to as few and simple ingredients as possible. For example, "Loose Peppers" is now only matched to "bell pepper"; \modi{see the middle right of Fig. \hspace{-.2em}\ref{fig:schem_data_struct}}. We then restrict to products only corresponding to one ingredient, and also remove products that are matched to unrepresentative generic ingredients (e.g. "vegetable"). We take a (generic) ingredient to be unrepresentative, if it shares less than half of its flavour compounds with the ingredients in the same category. As an example, if the product "Loose Aubergine" were only matched to the generic ingredient "vegetable" which shared less than half of its flavour compounds with all other vegetable ingredients, such as "asparagus", "lettuce" and "onion", we would exclude this product. This further reduces the number of products and ingredients of interest to \modi{be} $175$ and $69$ respectively, with $47222$ corresponding recipes, and each recipe being expected to contain $3$ such ingredients. 

\section{Methods}
\subsection{Product-purchase network}
\begin{figure}
    \centering
    \includegraphics[width=.8\textwidth]{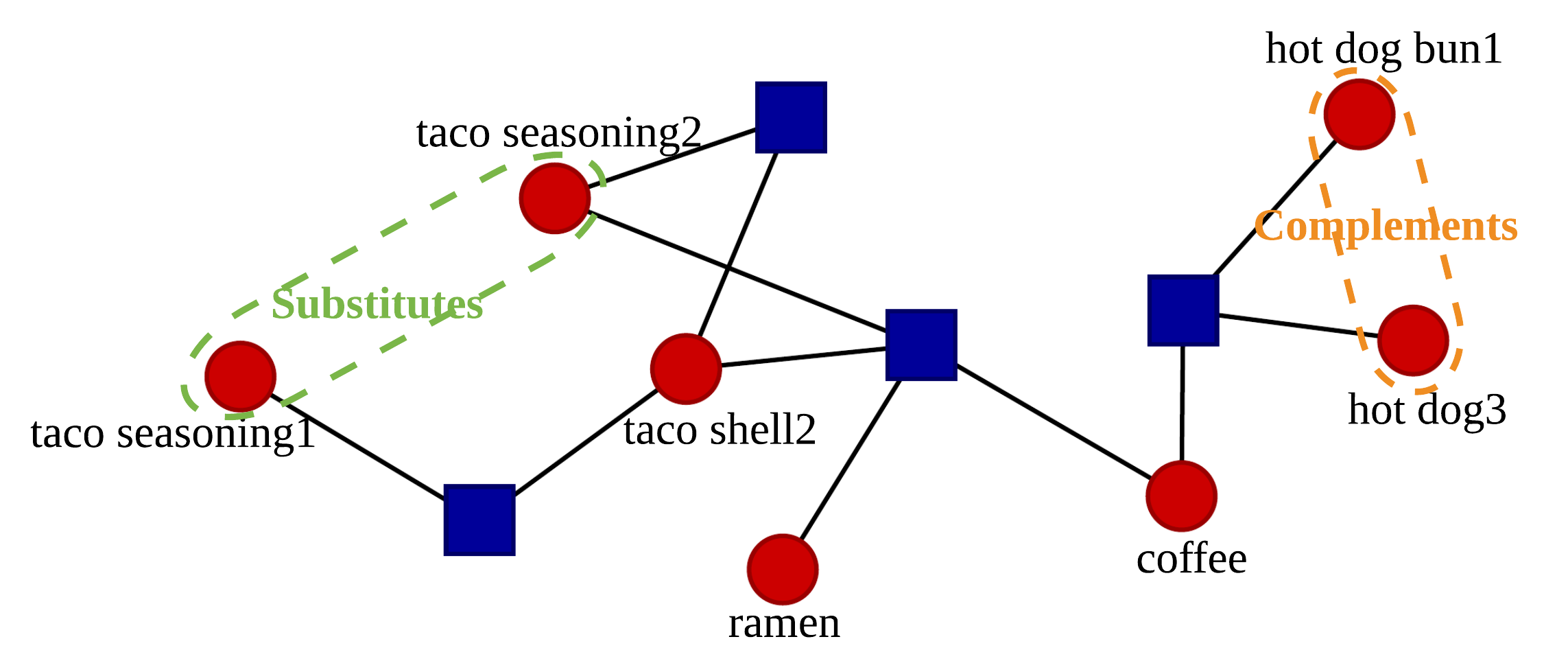}
    \caption{Example of a product-purchase network, where blue squares are transaction nodes, red circles are products nodes and these two sets of nodes are connected if the product is purchased in the corresponding transaction. The underlying sales data contains both complements (e.g. hot dog3 and hot dog bun1) and substitutes (e.g. taco seasoning1 and taco seasoning2).}
    \label{fig:eg_binet}
\end{figure}

We model the structure in the sales transaction data as a bipartite network, where we have two subsets of nodes, one corresponding to transactions and the other to products. A transaction node and a product node are connected, if the product is purchased in that particular transaction; see Figs. \hspace{-.2em}\ref{fig:schem_data_struct} (top) and \ref{fig:eg_binet}. We call it the \textit{product-purchase network}, and aim to extract product relationships from how product nodes are connected to each other in the network. This problem is generally related to the projection of bipartite networks to unipartite ones \cite{Zhou_bipartiteproj_2007}. Different strategies exist depending on the nature of the relationship that one wants to infer \cite{Zhou_bipartiteproj_2007, li_weighted_2005, newman_scientific_2001, newman_scientific-2_2001, newman_coauthorship_2004, leicht_vertexsim_2006}. While a majority of works look for assortative relations, in the sense that two nodes are connected in the unipartite network if they tend to share many neighbours in the bipartite one, more general types of projections can be defined, which are associated to the role played by the nodes in the \modi{bipartite} network, and are particularly relevant to extracting complements and substitutes. In the following section, we will specify our assumptions about the product relationships, which can be further interpreted as the \modi{specific} connectivity patterns in the product-purchase network; see typical examples in Fig. \hspace{-.2em}\ref{fig:eg_binet}.

We use the biadjacency matrix $\mathbf{A}^{(b)} = (A_{li})\in \{0,1\}^{n_t\times n_p}$ to represent the product-purchase network, where $n_t$ is the number of transaction nodes, $n_p$ is the number of product nodes, and $A_{li} = 1$ if product $i$ is purchased in transaction $l$ and $0$ otherwise. 

\subsection{\modi{Key} assumptions}
\label{sec:assumptions}
\modi{To characterise the product relationships, we consider the purchase patterns of products. Specifically,} in the context when prices change frequently, complements can be identified through sufficient co-purchases \cite{athey_comtheory_1998}, while substitutes have almost no co-purchases. The feature of substitutes that have similar interactions with other products is commonly used in practice \cite{ruiz_shopper_2019, chen_studying_2020}, and \modi{combined with} the almost-no-co-purchase characteristics, it can be used to determine the substitute relationship. \modi{Note that the formal definition through cross-price elasticity is expected to emerge from such purchase patterns, where, for example, two products always purchased together implies that the decrease in one’s price will result in an increase of the other’s demand\footnote{\footnotesize{The demand of a product is generally a decreasing function of its own price. This statement is true for the products analysed here.}}. Based on these arguments, we propose the following assumptions 1-4 to characterise complements and substitutes in the product-purchase network.}
\begin{enumerate}[label=\arabic*, font=\bfseries, before=\bfseries]
    \item Complements are products that are in the same transactions significantly more frequently than expected.
	\item The degree of complementarity between \modi{complements} is positively correlated with how frequently they are in the same transactions.
	\item Substitutes are products that share the same complements but are in the same transactions significantly less frequently.
	\item The degree of substitutability between \modi{substitutes} is positively correlated with how similar their complements are.
\end{enumerate}

\modi{In addition, we define \textit{noise} to be the purchase patterns that are caused by other, often unknown, factors and cannot be explained by complementarity and substitutability. Thus to capture the product relationships and their degrees, it is essential to control the noise effect. In networks, \textit{local structure} usually refers to the information around a node, and \textit{global structure} characterises the whole network. For intermediate scales, one often refers to the notion of \textit{mesoscale structure}, which is associated to groups of nodes that share similar connectivity patterns. Here, we consider particularly the \textit{community structure}, where groups of nodes are densely connected internally but sparsely connected externally. Within our context, we exploit the fact that the mesoscale structure is much more robust to noise than the local information \cite{donnat_scale_2018} Hence, we further propose the following assumptions 5 and 6 to restrict the noise effect. } 
\begin{enumerate}[label=\arabic*, font=\bfseries, before=\bfseries]
    \setcounter{enumi}{4}
	\item Noise will not change the \modi{community} structure of complements and substitutes, \modi{i.e. groups of products that are mostly complements and substitutes respectively}.
	\item \modi{Noise} can be explained by some random models so that \modi{its} effect on the \modi{local structure of the network} can be removed accordingly.
\end{enumerate}

\modi{Hence, in Sect. \hspace{-.2em}\ref{sec:null_models}, we will determine whether each pair of} products are complements or substitutes by applying significance tests on the number of common neighbours between each pair of nodes in the product-purchase network (i.e. the same transactions they are in, assumptions 1 and 3). This step corresponds to projecting the bipartite network on the product side to form two unweighted unipartite networks, showing the existence of the two relationships. Further, in Sect. \hspace{-.2em}\ref{sec:method_measures}, we will quantify the degrees of complementarity and substitutability by local measures based on the product nodes’ neighbourhood structure in the bipartite and projected networks, respectively (assumptions 2 and 4). This step further adds weights to the corresponding unipartite networks.

\subsection{Null models}
\label{sec:null_models}
We propose the following null models on the product-purchase network, to determine whether the number of common neighbours, $cn_{ij}$, between each pair of product nodes $i$ and $j$, is significantly more or significantly less, with significance levels $\alpha_m$ or $\alpha_l$, respectively. Accordingly, two unweighted unipartite networks only consisting of product nodes can be obtained: (i) $\mathbf{A}^{(m)} = (A^{(m)}_{ij}) \in \{0,1\}^{n_p\times n_p}$ where $A^{(m)}_{ij} = 1$ if and only if $cn_{ij}$ is significantly more; (ii) $\mathbf{A}^{(l)} = (A^{(l)}_{ij}) \in \{0,1\}^{n_p\times n_p}$ where $A^{(l)}_{ij} = 1$ if and only if $cn_{ij}$ is significantly less. Finally, by assumptions 1 and 3 in Sect. \hspace{-.2em}\ref{sec:assumptions}, two networks indicating the existence of product relationships can be constructed: (i) $\mathbf{A}^{(c)} = (A^{(c)}_{ij}) \in \{0,1\}^{n_p\times n_p}$ where $A^{(c)}_{ij} = 1$ if and only if products $i,j$ are complements; (ii) $\mathbf{A}^{(s)} = (A^{(s)}_{ij}) \in \{0,1\}^{n_p\times n_p}$ where $A^{(s)}_{ij} = 1$ if and only if products $i,j$ are substitutes.

\subsubsection{Variant of Bipartite Erd\H{o} s-R\'enyi (ER) Models} 
\label{sec:variant_biER}
The ER model assumes a fixed probability for each edge to appear, independently of the others \cite{erdos_renyi_1959}, while bipartite ER models only allow edges between the two subsets of nodes. 
	
In our variant, we assign a different connecting probability $p_{i}$ for each product $i$. Then, the probability that a transaction node is connected with both product nodes $i$ and $j$ is $p_{i}p_{j}$; the number of their common neighbours, $cn_{ij}$, is a random variable $X_{ij}$, s.t. $X_{ij} \sim B(n_t, p_{i}p_{j})$. We further assume $n_t$ is sufficiently large, and approximate the distribution by $N(\mu_{ij}, \sigma_{ij}^2)$, where $\mu_{ij} = n_tp_{i}p_{j}$, $\sigma_{ij}^2 = n_tp_{i}p_{j}(1-p_{i}p_{j})$, from the Central Limit Theorem \cite{Grimmett_probability_2001}. Hence, $cn_{ij}$ is significantly more if 
	\begin{align*}
	    cn_{ij} > \ n_tp_{i}p_{j} + \Phi^{-1}(1-\alpha_m)\sqrt{n_tp_{i}p_{j}(1-p_{i}p_{j})},
	\end{align*} 
	and is significantly less if 
	\begin{align*}
	    cn_{ij} < \ n_tp_{i}p_{j} - \Phi^{-1}(1-\alpha_l)\sqrt{n_tp_{i}p_{j}(1-p_{i}p_{j})},
	\end{align*} 
	where $\Phi^{-1}(\cdot)$ is the inverse cumulative function of $N(0,1)$, and the maximum likelihood estimate for each $p_i$ is
	\begin{align*}
	    \hat{p}_{i} = \frac{d_{i}^{(p)}}{n_t},
	\end{align*}
	where $d_i^{(p)}$ is the degree of product node $i$.
	
\subsubsection{Bipartite Configuration Models (BiCMs)}
\label{sec:bicm}
The configuration model creates a network with a given degree sequence $\{d_i\}$, by assigning $d_i$ half-edges (or stubs) to each node $i$ and joining two chosen stubs uniformly at random until no more stubs are left \cite{newman_networks_2018, newman_randomg_2001}. The BiCM takes the bipartite features into account, where two degree sequences are given, dividing the nodes into two subsets, and edges are only allowed between the two subsets of nodes. Note that multi-edges are allowed here, but since we assume finite variance in both degree distributions, they are negligible in large networks (see \modi{Appendix \ref{sec:app_bicm_edge_prob} for details}). 

The probability of product nodes $i,j$ sharing a transaction node $l$ is 
\begin{align*}
    p_{ilj} = \frac{d_i^{(p)}d_l^{(t)}d_j^{(p)}(d_l^{(t)}-1)}{m^2},
\end{align*} 
where the superscripts $t,p$ stand for transaction nodes and product nodes respectively, $d_h^{(\cdot)}$ is the degree of node $h$, and $m = \sum_{l=1}^{n_t}d_l^{(t)} = \sum_{i=1}^{n_p}d_i^{(p)}$ is the number of edges (see \modi{Appendix \ref{sec:app_bicm_edge_prob}} for details). The variant of bipartite ER models in Sect. \hspace{-.2em}\ref{sec:variant_biER} can be seen as an approximation of this model, where we assume that the degree of each transaction node is constant.  
	
The number of common neighbours between product nodes $i$ and $j$, $cn_{ij}$, is the sum of $Bernoulli(p_{ilj})$ over $l$, where $p_{ilj}$ \modi{possibly varies for different transaction node $l$}. We assume independence between different transaction nodes to connect with them both. Hence, $cn_{ij}$ is a Poisson binomial random variable, $X_{ij}$, with the mean value 
\begin{align*}
    \mu_{ij} = \sum_{l=1}^{n_t}\frac{d_i^{(p)}d_l^{(t)}d_j^{(p)}(d_l^{(t)}-1)}{m^2} = \frac{d_i^{(p)}d_j^{(p)}}{m}\frac{<d^{(t)2}> - <d^{(t)}>}{<d^{(t)}>},
\end{align*}
where $<d^{(t)}> = (\sum_{l=1}^{n_t}d_l^{(t)})/n_t$, and $<d^{(t)2}> = (\sum_{l=1}^{n_t}d_l^{(t)2})/n_t$.

\modi{The Poisson binomial distribution can be well approximated, with an exact error bound, by a Poisson distribution with the same mean, if the composing Bernoulli probabilities, $p_{ilj}$, are sufficiently small \cite{lecam_poisson_1960}. Since real networks are sparse, the $p_{ilj}$s are generally small (see our particular case in Appendix \ref{sec:app_poisson_approxi}). Hence, we use $Y_{ij}\sim Poisson(\mu_{ij})$ for the significance tests here, and determine} $cn_{ij}$ to be significantly more if
\begin{align*}
    \modi{1 - F_{ij}(cn_{ij}) < \alpha_m}, 
\end{align*}
and to be significantly less if
\begin{align*}
    \modi{F_{ij}(cn_{ij}) < \alpha_l},
\end{align*}
\modi{where $F_{ij}(y) = e^{-\mu_{ij}}\sum_{k=0}^{\floor*{y}}\mu_{ij}^k/k!$ is the cumulative distribution function of $Y_{ij}$.} 

\modi{The two null models are proposed to explain the purchase patterns purely from noise; with more information about the noise factors, one can propose more customised null models to explain more of such patterns. Currently, our null models are only based on difference in product popularity, and the BiCM also uses the heterogeneity in basket sizes: both are sufficiently general to incorporate additional noise factors, but could possibly not be sufficient in their current form, as hidden factors, e.g. correlated preference, could cause more common neighbours between product nodes in the product-purchase networks. Hence, by assumption 5 in Sect. \hspace{-.2em}\ref{sec:assumptions}, we accompany these null models with extra rules of significance-level selection: (i) $\alpha_m$ is chosen to be the smallest value that maintains the same community structure as that obtained from a baseline significance level, to exclude the above spurious signal; (ii) $\alpha_l$ is chosen to be the largest such value, in order not to accidentally filter out genuine patterns.}

\modi{Finally, we can obtain the unweighted network of complementary relationship, $\mathbf{A}^{(c)}$, and that of substitute relationship, $\mathbf{A}^{(s)}$.} By assumption 1, 
\begin{align*}
    \mathbf{A}^{(c)} = \mathbf{A}^{(m)};
\end{align*}
by assumption 3,
\begin{align*}
    \mathbf{A}^{(s)} =  \mathbf{I}_{\{\mathbf{A}^{(m)T}\mathbf{A}^{(m)} > 0\}}\odot\mathbf{A}^{(l)},
\end{align*}
where $\mathbf{I}_{\{\cdot\}}$ is the element-wise indicator matrix, and $\odot$ represents element-wise (Hadamard) matrix product.

\subsection{Measures}
\label{sec:method_measures}
The degrees of complementarity and substitutability matter. A significant relationship is not necessarily a strong relationship, and stronger relationships should be given higher weights to be more dominant in the networks. By assumption 2 in Sect. \hspace{-.2em}\ref{sec:assumptions}, the degree of complementarity is not directly correlated with how significant the co-purchase pattern is, but its relative frequency; by assumption 4 in Sect. \hspace{-.2em}\ref{sec:assumptions}, neither is the degree of substitutability, which causes the results in Sect. \hspace{-.2em}\ref{sec:null_models} not to be applicable here. Hence, in this section, we further propose measures to quantify both degrees, in order to convert the unweighted unipartite networks, $\mathbf{A}^{(c)}$ and $\mathbf{A}^{(s)}$, to weighted ones, $\mathbf{W}^{(c)}$ and $\mathbf{W}^{(s)}$, respectively, where $\mathbf{W}^{(c)}, \mathbf{W}^{(s)} \in [0,1]^{n_p\times n_p}$.

\subsubsection{Measures for complementarity}
We propose several measures for the degree of complementarity by interpreting assumption 2, where the more similar their neighbours in the product-purchase network are, the more complementary they are. 

\modi{We start from an \textit{enhanced version} of assumption 6 in Sect. \hspace{-.4em}\ref{sec:assumptions}: the noise factors change frequently and erratically so that their bias on the relative number of co-purchases between pairs of products can be neglected. We then propose the following measures,} derived from the weighted cosine similarity between random walkers starting from \modi{pairs} of nodes after one step. \modi{Specifically,} for each product node $i$, suppose that an impulse $\mathbf{y}_i(0) = \mathbf{e}_i\in \{0,1\}^{n_p}$, with value $1$ only in its $i$-th element, is injected on the product side at time $t=0$. We record the response of the system after a one-step random walk $\mathbf{y}_i(1) = \mathbf{P}^T\mathbf{y}_i(0)$, where $\mathbf{P} = \mathbf{D}^{(p)-1}\mathbf{A}^{(b)T}$, $\mathbf{A}^{(b)} = (A_{li})$ is the biadjacency matrix from the transaction nodes to the product nodes, and $\mathbf{D}^{(p)} = \mathbf{Diag}(d^{(p)}_i)$ is the diagonal matrix with the degrees of product nodes on its diagonal \cite{schaub_multiscale_2019}. We set the relative importance of each transaction $l$ as the inverse of its degree $d^{(t)}_l$, hence the weighted cosine similarity between \modi{the responses $\mathbf{y}_i(1)$ and $\mathbf{y}_j(1)$} is 
\begin{align*}
	sim(i,j) &= \frac{\mathbf{y}_i(1)^T\mathbf{W}_{cos}\mathbf{y}_j(1)}{||\mathbf{y}_i(1)||_{\mathbf{W}_{cos}}||\mathbf{y}_j(1)||_{\mathbf{W}_{cos}}} \\
	&= \sum_{l=1}^{n_t}\frac{\frac{A_{li}}{\sum_{k=1}^{n_t}A_{ki}}\frac{1}{d^{(t)}_l}\frac{A_{lj}}{\sum_{k=1}^{n_t}A_{kj}}}{\sqrt{(\sum_{h=1}^{n_t}\frac{A_{hi}}{\sum_{k=1}^{n_t}A_{ki}}\frac{1}{d^{(t)}_h}\frac{A_{hi}}{\sum_{k=1}^{n_t}A_{ki}})(\sum_{h=1}^{n_t}\frac{A_{hj}}{\sum_{k=1}^{n_t}A_{kj}}\frac{1}{d^{(t)}_h}\frac{A_{hj}}{\sum_{k=1}^{n_t}A_{kj}})}}\\
	&= \sum_{l=1}^{n_t}\frac{A_{li}A_{lj}}{d^{(p)}_id^{(t)}_ld^{(p)}_j\sqrt{(\sum_{h=1}^{n_t}\frac{A_{hi}}{d_i^{(p)2}d^{(t)}_h})(\sum_{h=1}^{n_t}\frac{A_{hj}}{d_j^{(p)2}d^{(t)}_h})}}\\
	&= \sum_{l=1}^{n_t}\frac{A_{li}A_{lj}}{d^{(t)}_l\sqrt{(\sum_{h=1}^{n_t}\frac{A_{hi}}{d^{(t)}_h})(\sum_{h=1}^{n_t}\frac{A_{hj}}{d^{(t)}_h})}},
\end{align*}
where $\mathbf{W}_{cos} = \mathbf{Diag}(1/d^{(t)}_l)$ is the weight matrix for the cosine similarity, and $||\mathbf{y}||_{\mathbf{W}} = \sqrt{\mathbf{y}^T\mathbf{W}\mathbf{y}} = ||\mathbf{W}^{1/2}\mathbf{y}||_2$ with $\mathbf{W}$ (symmetric) positive-definite. This \modi{introduces the first measures we propose, the} \textit{original measure},
\begin{align}
	sim_{o}(i,j)= \sum_{l=1}^{n_t}\frac{A_{li}A_{lj}}{d^{(t)}_l\sqrt{(\sum_{h=1}^{n_t}\frac{A_{hi}}{d^{(t)}_h})(\sum_{h=1}^{n_t}\frac{A_{hj}}{d^{(t)}_h})}},
	\label{equ:sim_o}
\end{align}
where $\mathbf{A}^{(b)} = (A_{li})$, $\{d^{(t)}_l\}_{l=1}^{n_t}$ and $n_t$ are the same as before. Hence, each common neighbour $A_{li}A_{lj}$ between each pair of product nodes $i,j$ \modi{in the product-purchase network} is discounted by the degree of the corresponding transaction node $l$, and this quantity is further scaled so that each product is at the maximum level of complementarity to itself, i.e value $1$, in a symmetric manner. A higher value means relatively more common neighbours of lower degrees. Naturally, we also propose the \textit{original directed measure}, 
\begin{align}
    sim_{od}(i,j)= \sum_{l=1}^{n_t}\frac{A_{li}A_{lj}}{d^{(t)}_l(\sum_{h=1}^{n_t}\frac{A_{hj}}{d^{(t)}_h})},
\label{equ:sim_od}
\end{align}
where each $(i,j)$ entry measures the degree of complementarity of product $i$ to product $j$. Compared with those in the literature, our measures are globally comparable, where node pairs with no common node can also be compared.

\modi{The above enhanced version of assumption 6 is reasonable for our choice of fresh food, since the price has been changed frequently and erratically, as required, during the chosen time period. For a general product, in contrast, it would be necessary to implement assumption 6, in order to properly remove the noise effect from} our measures. However, most literature followed the direction of filtering out insignificant edges, rather than removing noise from network measures. In this article, we take an initial step in the latter direction by deducting the mean value with some noise models.

First, we should determine which quantity to subtract the mean from. If we consider the original measure as the geometric mean, 
\begin{align*}
	sim_{o}(i,j)= \sqrt{\frac{\sum_{l=1}^{n_t}\frac{A_{li}A_{lj}}{d^{(t)}_l}}{\sum_{h=1}^{n_t}\frac{A_{hi}}{d^{(t)}_h}}\frac{\sum_{l=1}^{n_t}\frac{A_{li}A_{lj}}{d^{(t)}_l}}{\sum_{h=1}^{n_t}\frac{A_{hj}}{d^{(t)}_h}}} = \sqrt{\frac{\sum_{l\in\Gamma(i)\cap \Gamma(j)}\frac{A_{li}}{d^{(t)}_l}}{\sum_{h\in\Gamma(i)}\frac{A_{hi}}{d^{(t)}_h}}\frac{\sum_{l\in\Gamma(j)\cap\Gamma(i)}\frac{A_{lj}}{d^{(t)}_l}}{\sum_{h\in\Gamma(j)}\frac{A_{hj}}{d^{(t)}_h}}},
\end{align*} 
where $\Gamma(i) = \{l: A_{li} = 1\}$ is the set of node $i$'s neighbours in the product-purchase network, then we can propose the corresponding \textit{randomised measure},
\begin{align}
		sim_{r}(i,j)=  \sqrt{\frac{\sum_{l\in\Gamma(i)\cap \Gamma(j)}\left(\frac{A_{li}}{d^{(t)}_l}\ - \mathbb{E}\left[\frac{A^{(r)}_{li}}{d^{(r)}_l}\right]\right)}{\sum_{h\in\Gamma(i)}\left(\frac{A_{hi}}{d^{(t)}_h} - \mathbb{E}\left[\frac{A^{(r)}_{hi}}{d^{(r)}_h}\right]\right)}\frac{\sum_{l\in\Gamma(j)\cap \Gamma(i)}\left(\frac{A_{lj}}{d^{(t)}_l} - \mathbb{E}\left[\frac{A^{(r)}_{lj}}{d^{(r)}_l}\right]\right)}{\sum_{h\in\Gamma(j)}\left(\frac{A_{hj}}{d^{(t)}_h} - \mathbb{E}\left[\frac{A^{(r)}_{hj}}{d^{(r)}_h}\right]\right)}},
		\label{equ:sim_r}
\end{align}
where $\mathbf{A}^{(b)} = (A_{li})$, $\mathbf{A}^{(r)} = (A^{(r)}_{li})$ is the corresponding biadjacency matrix of a random product-purchase network with each $A^{(r)}_{li}$ being a random variable, and $d^{(r)}_{l} = \sum_{i=1}^{n_p}A^{(r)}_{li}$. 
The \textit{randomised directed measure} naturally follows to be 
\begin{align}
		sim_{rd}(i,j)=  \frac{\sum_{l\in\Gamma(j)\cap \Gamma(i)}\left(\frac{A_{lj}}{d^{(t)}_l} - \mathbb{E}\left[\frac{A^{(r)}_{lj}}{d^{(r)}_l}\right]\right)}{\sum_{h\in\Gamma(j)}\left(\frac{A_{hj}}{d^{(t)}_h} - \mathbb{E}\left[\frac{A^{(r)}_{hj}}{d^{(r)}_h}\right]\right)}.
		\label{equ:sim_rd}
\end{align}

Next, we should determine the noise model. For example, assuming fixed basket sizes (transaction node degrees) and product purchase frequencies (product node degrees), naturally leads us to \modi{the} BiCM (cf. Sect. \hspace{-.2em}\ref{sec:null_models}) as our noise model. In this particular case, 
\begin{align}
	\mathbb{E}\left[\frac{A^{(r)}_{li}}{d^{(r)}_l}\right] = \frac{d^{(p)}_i}{m},
	\label{equ:measure bicm cn}
\end{align}
where $d^{(p)}_i$ is the degree of product node $i$, and $m$ is the number of edges in the product-purchase network. With Equations (\ref{equ:sim_r}), (\ref{equ:sim_rd}) and (\ref{equ:measure bicm cn}), we accordingly introduce the \textit{randomised configuration measure} and the \textit{randomised configuration directed measure}. 

\modi{Since the measures can be computed for any product pairs, but by assumption 2, only product pairs with value $1$ in $\mathbf{A}^{(c)}$ can be assigned positive degrees. Hence, the weighted adjacency matrix of complement unipartite network is obtained by}
\begin{align*}
    \mathbf{W}^{(c)} = \mathbf{A}^{(c)}\odot\mathbf{sim}_{\dagger},
\end{align*} 
where the subscript $\dagger$ can be $o$, $r$, $od$ or $rd$, \modi{and $\mathbf{sim}_{\dagger} = (sim_{\dagger}(i,j))\in [0,1]^{n_p\times n_p}$}. We call the values in $\mathbf{W}^{(c)}$ the \textit{complementarity scores}, and determine a pair of products to be complements if they have a positive complementarity score. 

\subsubsection{Measures for substitutability}
We propose measures for the degree of substitutability by assumption 4, where the more similar their complements are, the more substitutable they are. \modi{Here, we characterise each product by a vector of its complementarity scores with the other products, and use the (unweighted) cosine similarity between these vectors to indicate the degree of substitutability between pairs of products}. Specifically, for \modi{a} pair of nodes $i,j$,
\begin{align}
	sim_{s}(i,j) = \sum_{k=1}^{n_p}\frac{W_{ik}^{(c)}W_{jk}^{(c)}}{\sqrt{(\sum_{p=1}^{n_p}W_{ip}^{(c)2})(\sum_{p=1}^{n_p}W_{jp}^{(c)2})}},
	\label{equ:sim_s}
\end{align}
where $\mathbf{W}^{(c)} = (W_{ij}^{(c)})$ is the weighted adjacency matrix of complement unipartite network, and $n_p$ is the number of products. The substitutability measures are named after the complementarity measure used in $\mathbf{W}^{(c)}$. For example, with the original measure, we have the \textit{original substitutability measure}; with the randomised configuration measure, we have  the \textit{randomised configuration substitutability measure}. Naturally, we also propose the directed version, where for \modi{a} pair of nodes $i,j$,
\begin{align}
	sim_{sd}(i,j) = \sum_{k=1}^{n_p}\frac{\modi{\min(W_{ik}^{(c)}, W_{jk}^{(c)})}W_{jk}^{(c)}}{\sum_{p=1}^{n_p}W_{jp}^{(c)2}},
	\label{equ:sim_sd}
\end{align}
\modi{where the minimum function is used to guarantee that the measure reaches its maximum value when the complementarity degrees of product $i$ to others are no less than the corresponding degrees of product $j$.}

\modi{Since these measures can also be computed for any product pairs, but by assumption 4, only product pairs of value $1$ in $\mathbf{A}^{(s)}$ can be assigned positive degrees. Hence the weighted adjacency matrix of substitute unipartite network is obtained by}
\begin{align*}
    \mathbf{W}^{(s)} = \mathbf{A}^{(s)}\odot\mathbf{sim}_{\dagger},
\end{align*}
where the subscript $\dagger$ stands for $s$ or $sd$, \modi{and $\mathbf{sim}_{\dagger} = (sim_{\dagger}(i,j))\in [0,1]^{n_p\times n_p}$}. We name the values in $\mathbf{W}^{(s)}$ the \textit{substitutability scores}, and define a pair of products to be substitutes if they have a positive substitutability score. 

Note the measures of substitutability are based on those of complementarity \modi{and we do not apply extra noise removing strategies here, thus} it is critical that the complementarity degree is thresholded appropriately so that the substitutability degree is not biased by low-complementarity-degree products. 
\modi{Hence, by assumption 5 in Sect. \hspace{-.4em}\ref{sec:assumptions}, we accompany these measures with the following rules of threshold selection in analysing real data: (i) the threshold of the complementarity measures, $\theta_c$, is chosen to be the largest value that maintain the same community structure as that obtained from a baseline threshold value; (ii) the threshold of the substitutability measures, $\theta_s$, is chosen to be the smallest such value, for general noise removing purpose.}

\subsection{Role extraction}
Since both the null models in Sect. \hspace{-.4em}\ref{sec:null_models} and the measures in Sect. \hspace{-.4em}\ref{sec:method_measures} are based on local patterns in the product-purchase network directly or indirectly, so are the complement unipartite network, $\mathbf{W}^{(c)}$, and the substitute unipartite network, $\mathbf{W}^{(s)}$. It is then interesting to go beyond local patterns and explore the features between the node level and the whole network, the mesoscale structure, in such networks, i.e. groups of complements and groups of substitutes. 

One important type of mesoscale feature is the community structure, as in Sect. \hspace{-.4em}\ref{sec:assumptions}, where communities are groups of nodes that are densely connected internally but sparsely connected externally \cite{porter_communities_2009, fortunato_cdg_2010, fortunato_cdn_2016}. Various algorithms exist by virtue of interdisciplinary expertise \cite{newman_cdeig_2006, traag_leiden_2019, lambiotte_markpvst_2014, rosvall_map_2009, peixoto_bayesian_2019}, \modi{generally aiming to optimise a quality function with respect to different partitions of the network}. Here, we choose the information-theoretic (hierarchical) map equation \cite{rosvall_map_2009}, which aims to describe the trajectory of random walkers on the network most efficiently, thereby capturing the right community structure of the underlying network, and is known for being not affected by a common problem of community detection algorithms, the resolution limit \cite{kawamoto_estimating_2015}. From the detected structure, we will also examine the underlying assumption that groups are clique-like.

Considering the problem of extracting these two kinds of \modi{product} groups in the bipartite product-purchase network, it corresponds to a more general problem, \textit{role extraction}. \textit{Roles} are general versions of communities, where nodes inside the same role share similar connectivity patterns across the network \cite{lorrain_structural_1971, white_graph_1983, holland_expofam_1981}. Hence, it contains both classic \textit{assortative communities}, as described before, and \textit{disassortative communities}, where nodes are loosely connected internally while densely connected externally. We define the \textit{role adjacency} as $\mathbf{B} = (B_{rs})$, where
\begin{align*}
	B_{rs} = \sum_{i\in \mathcal{C}^r,\ j\in \mathcal{C}^s}\frac{W_{ij}}{n^{(r)}n^{(s)}},
\end{align*} 
$\mathcal{C}^r$, $r = 1,2,\dots$, are the roles, $n^{(r)} = |\mathcal{C}^r|$ for each role $r$, and $\mathbf{W} = (W_{ij})$ is the (weighted) adjacency matrix of the underlying network. The matrix \modi{$\mathbf{B}$} is induced by the maximum-likelihood estimate of the expected weights between nodes \modi{inside} the corresponding role(s) in the standard stochastic block model \cite{karrer_sbm-cd_2011}. Then, $\mathcal{C}^r$ is an assortative community if $B_{rs} \ll B_{rr},\ \forall s\ne r$; $\mathcal{C}^r$ is a disassortative community if $B_{rs} \gg B_{rr},\ \exists s$, \modi{i.e. community $r$ is much more densely connected with at least one other community $s$ than itself}. Thus, our set of methods establishes an indirect solution to the role extraction in bipartite networks. We call our detected groups of complements, the  \textit{complement roles}, and our detected groups of substitutes, the \textit{substitute roles}.

\subsection{External validation: product hierarchy, flavour compound and recipe data}
We start from the product hierarchy information to characterise both complement roles and substitute roles, and then check if the characteristics are consistent with the common understanding of complements and substitutes. Specifically, we exclusively use the \texttt{L3} product hierarchy, consisting of fruit (F), organic produce (OP), prepared produce (PP), salad (S), and vegetable (V). 

Next, we use the correspondence between flavour compounds and products to compute the Jaccard index, i.e. the relative number of shared flavour compounds, $rf_{ij}$, between each pair of products $i$ and $j$, 
\begin{align*}
    rf_{ij} = \frac{|C(i) \cap C(j)|}{|C(i) \cup C(j)|},
\end{align*}
where $C(i)$ is the set of all flavour compounds in product i. We then consider the cases in which $rf_{ij} = 0$ and $rf_{ij} = 1$, and check if the complementary pairs have a higher probability to share no flavour compounds and if the substitute pairs have a higher probability to share all their flavour compounds. Furthermore, we examine the relationship between $rf_{ij}$ and $W_{ij}^{(c)}, W_{ij}^{(s)}$, in terms of the \textit{Pearson} correlation, as well as the \textit{Spearman} correlation. 

Subsequently, we use the recipe data to evaluate the relative number of shared recipes, $rr_{ij}$, between each pair of products $i$ and $j$,
\begin{align*}
    rr_{ij} = \frac{|R(i) \cap R(j)|}{ |R(i) \cup R(j)|},
\end{align*}
where $R(i)$ is the set of all recipes including product i, and we set $rr_{ij} = 0$ if products $i$ and $j$ are matched to the same ingredient. We then assess if the complementary pairs and substitute pairs have significantly higher and lower probabilities to co-appear in relatively more recipes, respectively. This is achieved by the \textit{Mann-Whitney-Wilcoxon} (MWW) tests, where 
\begin{align*}
    \begin{cases}
       &H_0: P(X > Y) = P(Y > X),\\
       &H_1: P(X > Y) \ne or > or < P(Y > X),
    \end{cases}
\end{align*}
and $X$, $Y$ are two independent random variables \cite{mann_two_variable_test, fay_wmw-t_2010}. For example, let $X$ be the relative number of shared recipes from all product pairs $\{rr_{ij}\}$, and $Y$ be that from only complementary pairs $\{rr_{ij}: W_{ij}^{(c)} > 0\}$. Then, we will use the alternative hypothesis $H_1: P(X > Y) < P(Y > X)$. Similarly, we also explore the relationship between $rr_{ij}$ and $W_{ij}^{(c)}, W_{ij}^{(s)}$. 

\begin{figure}
    \centering
    \begin{subfigure}{.4\textwidth}
        \centering
        \includegraphics[width=.9\textwidth]{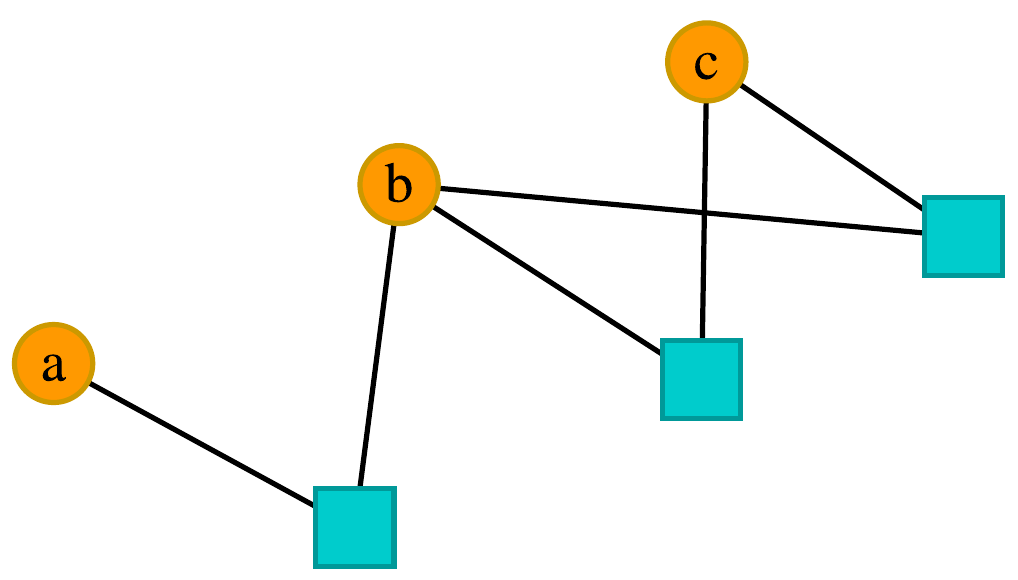}
        \caption{An ingredient-recipe network.}
    \end{subfigure}
    \begin{subfigure}{.25\textwidth}
        \centering
        \includegraphics[width=.9\textwidth]{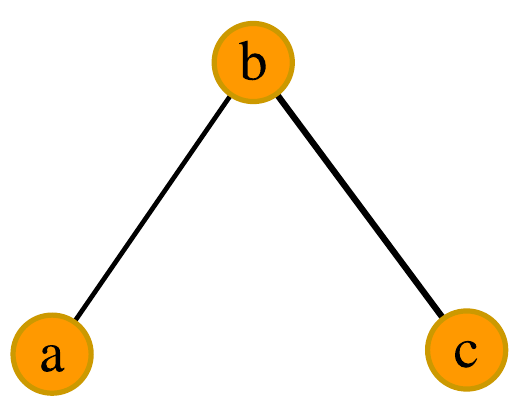}
        \caption{The weighted complement network of ingredients.}
    \end{subfigure}
    \begin{subfigure}{.25\textwidth}
        \centering
        \includegraphics[width=.9\textwidth]{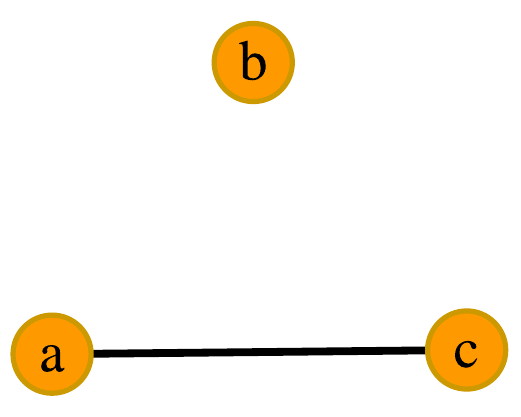}
        \caption{The weighted substitute network of ingredients.}
    \end{subfigure}
    \begin{subfigure}{.4\textwidth}
        \centering
        \includegraphics[width=.9\textwidth]{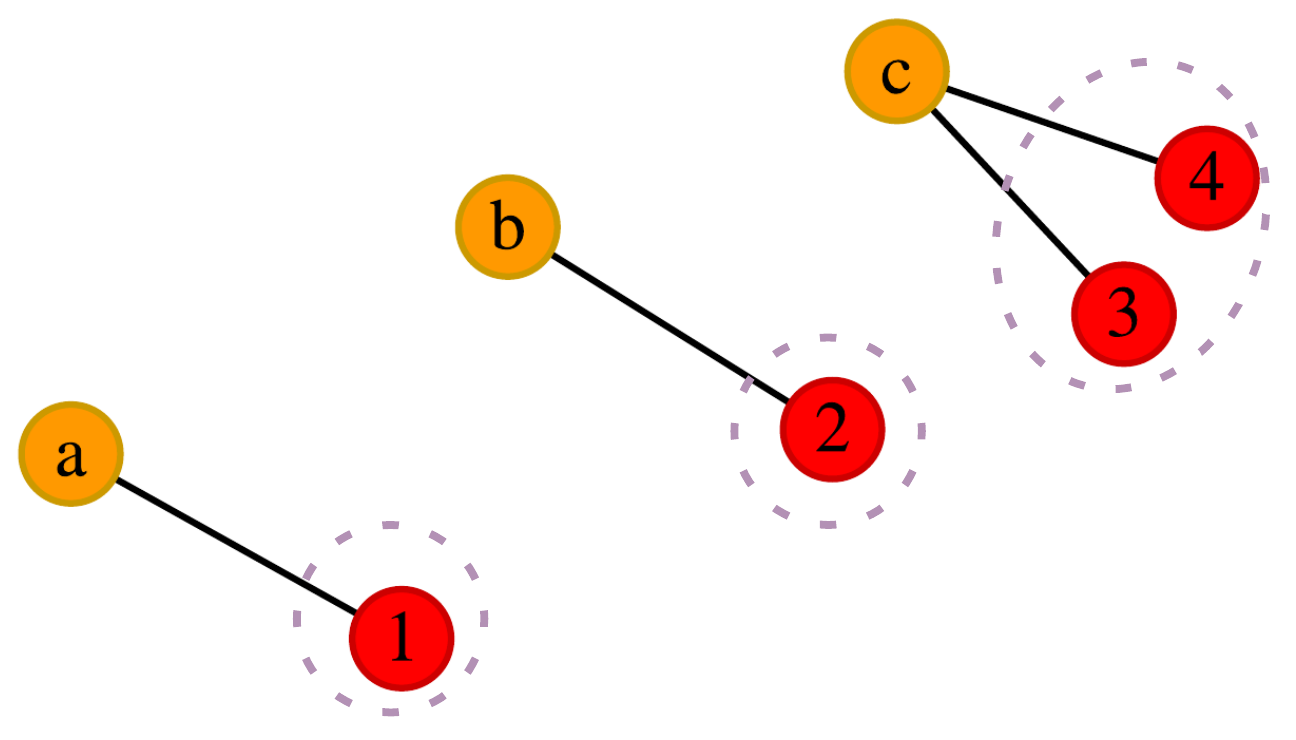}
        \caption{The matching between ingredients and products.}
    \end{subfigure}
    \begin{subfigure}{.25\textwidth}
        \centering
        \includegraphics[width=.95\textwidth]{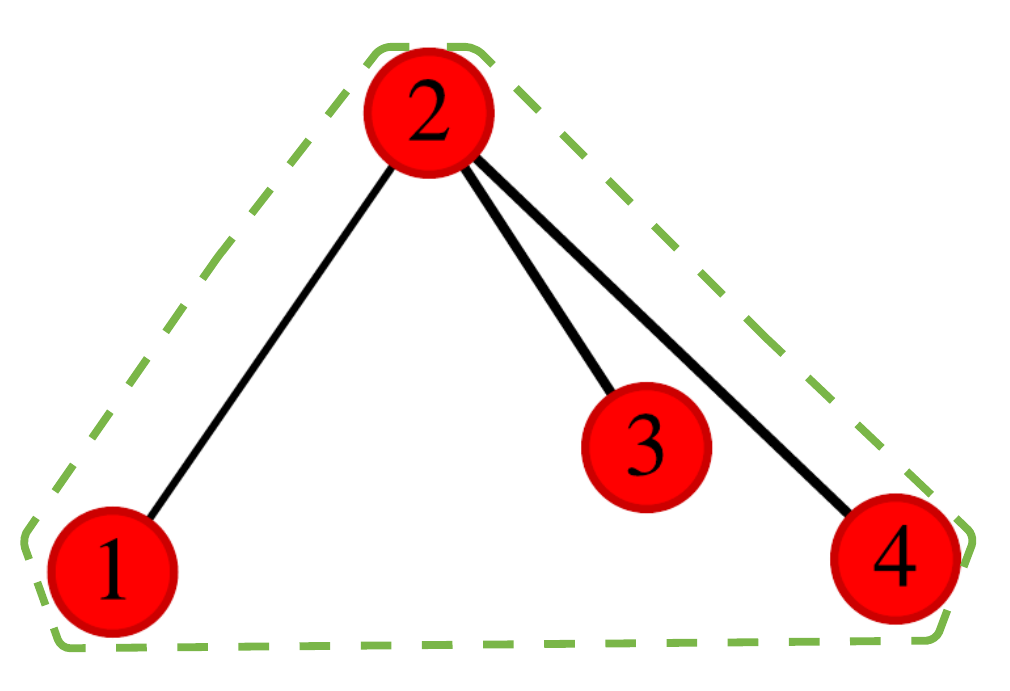}
        \caption{$\mathbf{W}^{(cr)}$.}
    \end{subfigure}
    \begin{subfigure}{.25\textwidth}
        \centering
        \includegraphics[width=.87\textwidth]{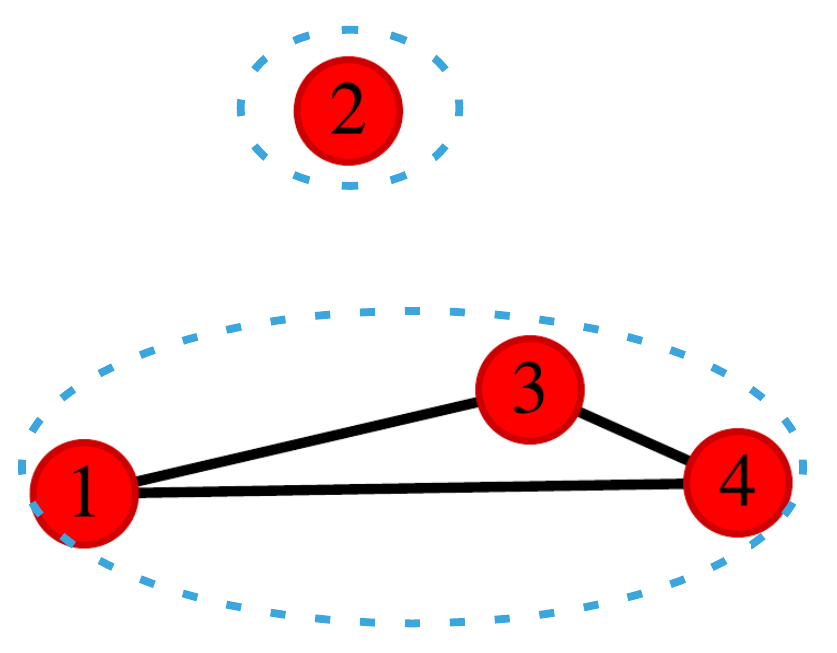}
        \caption{$\mathbf{W}^{(sr)}$.}
    \end{subfigure}
    \caption{Illustration of the process to compute the product roles from the recipe data, where cyan squares are recipe nodes, orange circles are ingredient nodes, red circles are product nodes, the line thickness corresponds to how high the corresponding scores are, and $l_0$ substitute roles, $l$ complement role(s) and $l_1$ substitute roles are shown as groups of product nodes in the purple dashed circles, green dashed polygon(s) and blue dashed circles, respectively.}
    \label{fig:illus_role}
\end{figure}

Finally, we apply our overall framework to the recipe data, where we treat recipes as transactions and ingredients as products. This stems from the hypothesis that customers purchase products to cook dishes following recipes, and thus the recipe data should be a restriction of the sales data. We compare the values of complementarity scores by recipes, the \textit{recipe complementarity scores} $\mathbf{W}^{(cr)} = (W_{ij}^{(cr)})$, with those by sales, $\mathbf{W}^{(c)}$, and similarly, the \textit{recipe substitutability scores} $\mathbf{W}^{(sr)} = (W_{ij}^{(sr)})$, with $\mathbf{W}^{(s)}$. Note that we set $W_{ij}^{(cr)} = 0$ and $W_{ij}^{(sr)} = 1$ if product $i$ and $j$ are matched to the same ingredient. We finish the validation stage by comparing the role assignments (of products) from both datasets, where $l$ complement roles and $l_1$ substitute roles (from the recipe data) are obtained from applying community detection on $\mathbf{W}^{(cr)}$ and $\mathbf{W}^{(sr)}$, respectively. We construct extra $l_0$ substitute roles by grouping together products that are matched to the same ingredients, for reference; see Fig. \hspace{-.2em}\ref{fig:illus_role} for details.

\section{Results}
\subsection{Illustrative example}
\label{sec:illustrative_example}
Before investigating noisy real data, we first validate our overall framework in a controlled "ideal world" where the relationship between products is known. Specifically, we simulate a consumer population characterised by a set of rules in this world, and ask whether our null models capture the right relationship between each pair of products, whether our measures give the right degree between them, and finally, whether our complement and substitute roles provide insights into the groups of complements, and the groups of substitutes, respectively. 

\begin{figure}
    \centering
    \includegraphics[width=.95\textwidth]{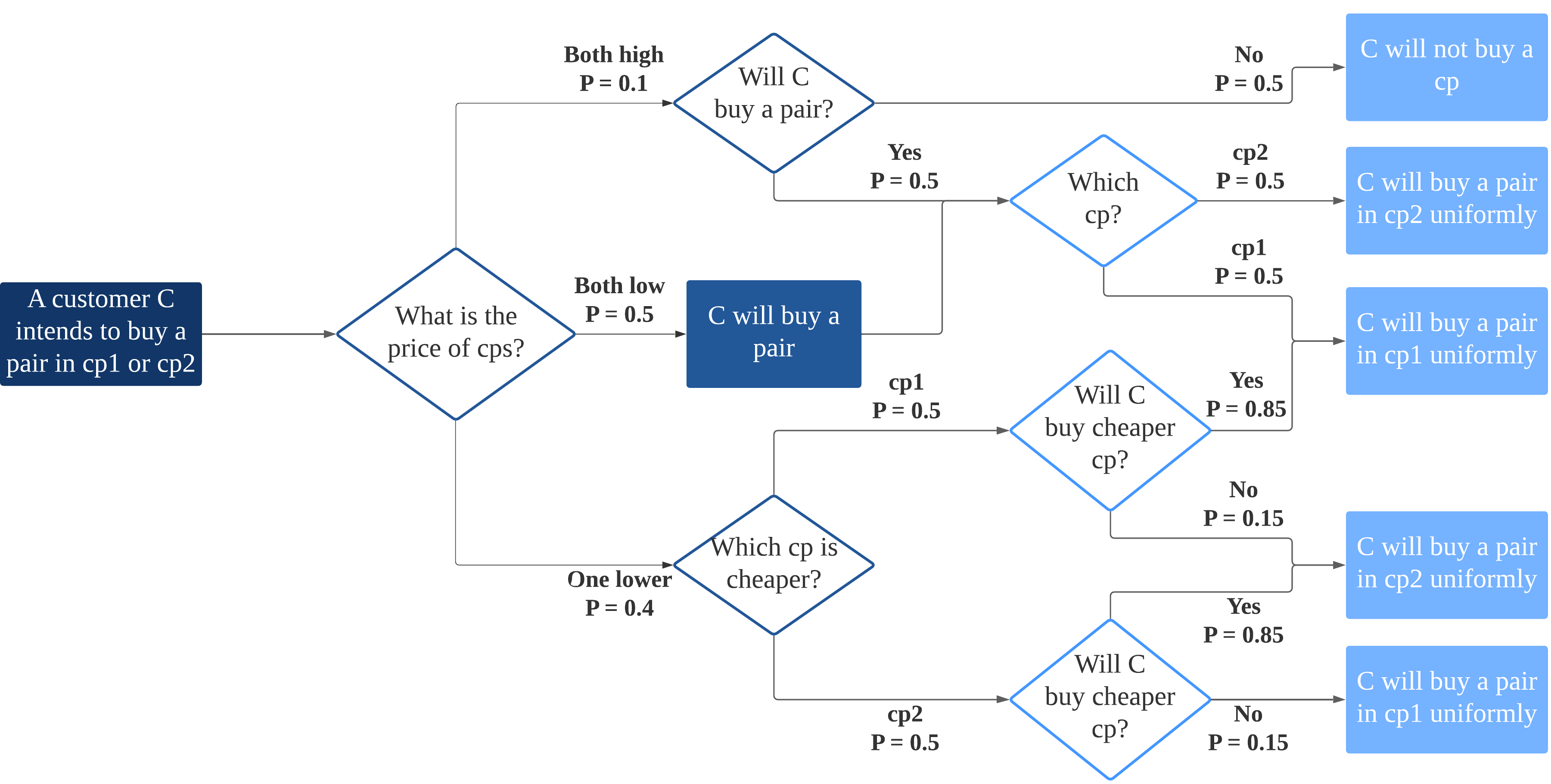}
    \caption{Illustration of how each customer chooses a complementary pair (cp), where cp1 and cp2 correspond to the \textit{hot-dog}-and-\textit{hot-dog-bun} and \textit{taco-shell}-and-\textit{taco-seasoning} complementary pairs, respectively.}
    \label{fig:illustr_buy_diagram}
\end{figure}

The simulated world is summarised as follows, similar to the one in \cite{ruiz_shopper_2019}. 
\begin{itemize}
	\item There are $13$ different products: \textit{coffee, wipes, ramen, candy, hot dog1, hot dog2, hot dog3, hot dog bun1, hot dog bun2, taco shell1, taco shell2, taco seasoning1, taco seasoning2}.
	
	\item \textit{coffee, wipes, ramen, candy} are independent products, but are popular with the customers, so are bought frequently. This corresponds to one possible source of noise, correlated preference, where the items are preferred by some customers but purchase decisions are made independently from one another, based on their features, e.g. price. 
	
	\item The other products form substitute groups and complementary pairs. Products of the same names ignoring the number at the end are groups of substitutes; pairs in \{\textit{hot dog1, hot dog2, hot dog3}\}$\times$\{\textit{hot dog bun1, hot dog bun2}\} and \{\textit{taco shell1, taco shell2}\}$\times$\{\textit{taco seasoning1, taco seasoning2}\} are complementary pairs. In this world, customers never buy just one item in a complementary pair, and they always buy at most one of all such pairs.
	
	\item Customers are sensitive to price. When the price of a popular product is low, they buy it with probability $0.8$; otherwise, they buy it with probability $0.2$. Each customer purchases each preferred product independently. 
	
	Sensitivity to the price of complementary pairs is different, since the probability to purchase a pair will decrease even if only one item in the pair has a high price. Hence, each pair is treated as a whole here. When all complementary pairs are of low price, customers buy one of them evenly; the case when all pairs are of high price is similar, except that customers have a $0.5$ chance not to buy any of them; when one of the pairs has a lower price than the others, they buy this one with probability $0.85$, and have $0.15$ probability to buy others evenly; see Fig. \hspace{-.2em}\ref{fig:illustr_buy_diagram} for details.
\end{itemize}
With these specifications, we simulate $1000$ transactions from this customer population. For a single transaction, each independent product has an $80\%$ chance of being marked up to a high price; there is a $50\%$ chance that all complementary pairs are of low price, a $10\%$ chance that all are of high price, and accordingly a $40\%$ chance that some are marked up, where the lowest priced one is chosen uniformly at random.

\begin{figure}
	\centering
	\includegraphics[width=.95\textwidth]{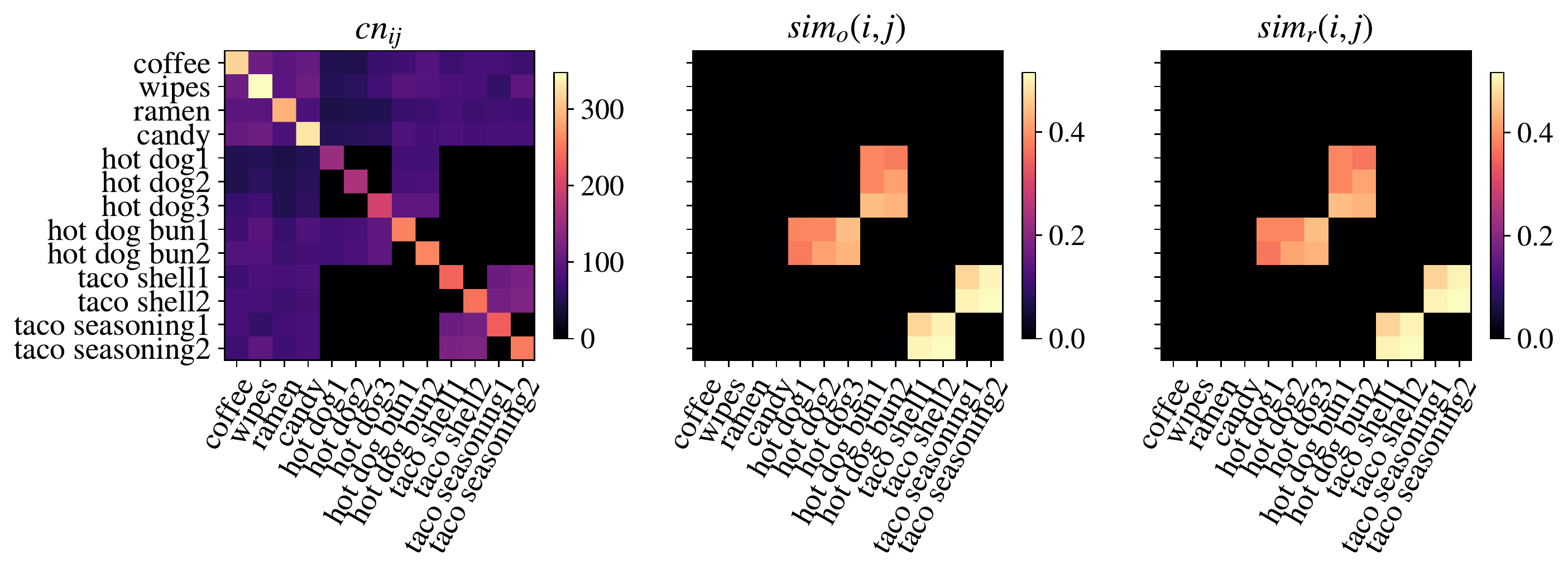}
	\caption{Measures on the products, from co-purchases $(cn_{ij})$ \modi{whose diagonal shows the purchase frequency} (\textbf{left}), the complementarity scores $\mathbf{W}^{(c)}$ induced by the original measure $(sim_o(i,j))$ (\textbf{middle}) and by the randomised configuration measure $(sim_r(i,j))$ (\textbf{right}), where x-axis, y-axis are products \modi{in the same order as being} listed in the simulated world assumptions.}
	\label{fig: demon sims}
\end{figure}

We provide the complementarity scores, $\mathbf{W}^{(c)}$, induced by the original measure, $\mathbf{sim}_o$, and by the randomised configuration measure, $\mathbf{sim}_r$, together with the number of co-purchases, $(cn_{ij})$, in Fig. \hspace{-.31em}\ref{fig: demon sims}. We choose the variant of ER model as the underlying null model, since it better explains the noise here\footnote{\footnotesize{Hence, the significance level for the significantly more co-purchases, $\alpha_c$, for the variant of ER model can be chosen as high as $0.9$ with the same results. While for the BiCM, a much lower significance level (e.g. $10^{-4}$ for $\alpha_c$) is needed in order to extract the true product relationships.}}. Note that independent products are bought more frequently, and their numbers of co-purchases with other products are fairly similar to those \modi{within} complementary pairs. However, our extracted complementary pairs $\{(i,j): W_{ij}^{(c)} > 0\}$ successfully retrieve the ground-truth complementary pairs. Accordingly, our extracted substitute pairs $\{(i,j): W_{ij}^{(s)} > 0\}$ successfully retrieve the ground-truth substitute pairs. Furthermore, the complementarity scores of the \textit{hot-dog}-and-\textit{hot-dog-bun} complementary pairs is between $0.3$ and $0.5$, and those of the \textit{taco-shell}-and-\textit{taco-seasoning} complementary pairs is around $0.5$. These values are approaching the inverse of the number of products in the corresponding substitute groups, which is consistent with the assuming complete substitution. 

\begin{figure}
	\centering
	\includegraphics[width=.8\textwidth]{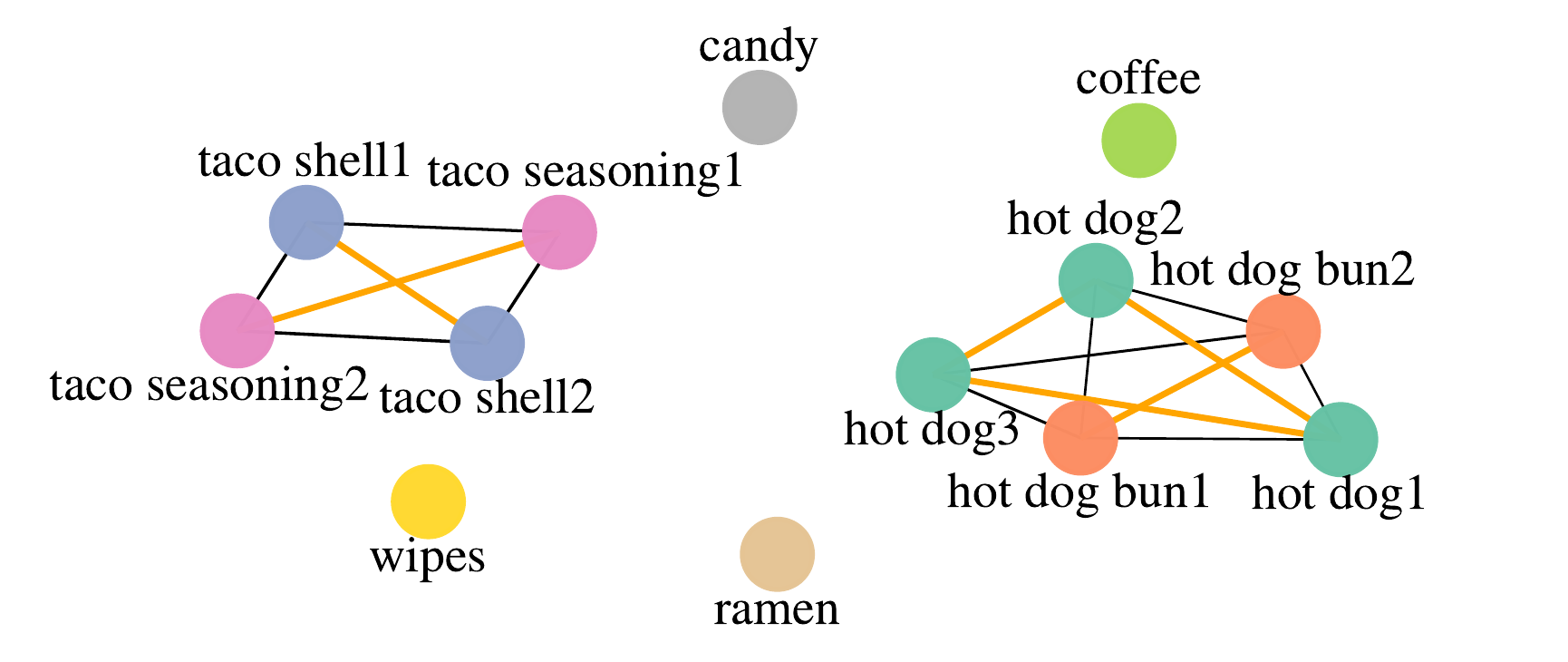}
	\caption{The unipartite network in the illustrative example, with product nodes connected by both the complementarity scores $(W_{ij}^{(c)})$ (in black) and the substitutability scores $(W_{ij}^{(s)})$ (in orange) induced by the original measure, where the line thickness corresponds to how high the scores are, and products in the same substitute role are shown in the same colour.}
	\label{fig: demon res}
\end{figure}

Finally, our substitute roles exactly agree with the ground-truth substitute groups; see Fig. \hspace{-.2em}\ref{fig: demon res}. Our complement roles reproduce the ground-truth complementary pairs including their corresponding groups of substitutes. Note that there are no groups of complements beyond the pairwise relationship. 

This example demonstrates the ability of our overall framework to determine both product relationships and their corresponding degrees, which paves the way for us to continue the analysis on real-world data. From a mesoscale perspective, our complement roles and substitute roles have much overlap with the groups of complements and those of substitutes, respectively. Furthermore, the fact that we already have complement roles involving substitutes indicates that the interaction between the two relationships is not negligible. For instance, it is entirely possible that we may find substitute roles including complements in real data.

\subsection{Sales data}
\label{sec:results_salesdata}
\modi{Hereafter, we use the variant of ER model as the underlying null model, since its assumptions are generally applicable in real-world purchases, and we only show the results from the original measure, because both have very similar behaviour; see Appendix \ref{sec:app_sales_data} for the parameter calibration and the results from the randomised measure.} We \modi{first} examine the ranking power of our scores, $\mathbf{W}^{(c)}$ and $\mathbf{W}^{(s)}$, by checking the top complementary pairs and substitute pairs for each product. This is done by choosing several query products $j$ at random, and output the products of the three highest complementarity scores $W_{ij}^{(c)}$ and the ones of the three highest substitutability scores $W_{ij}^{(s)}$; see Table \ref{tab:rank} for one run. The substitute pairs of scores $>0.1$ largely agree with common sense\footnote{\footnotesize{Here we refer to the notion that products that are essentially the same are substitutes, for example, Brand A apples and Brand B apples.}}. For example in Table \ref{tab:rank}, the top substitute of organic blueberries is blueberries, and the top substitutes of salad tomatoes are other types of tomatoes. Additionally, the ranking indicates that common-sense substitutes have high complementarity scores with the same products. For example salad tomatoes, baby plum tomatoes and tomatoes on the vine are the top three complements of loose cucumbers. These findings justify our assumption 3 in Sect. \hspace{-.4em}\ref{sec:assumptions}. There are also some nontrivial substitutes, of lower score values, from general understanding, which we will discuss in Sect. \hspace{-.2em}\ref{sec:discussion}. 
\begin{table}
    \centering 
	\caption{Products of the three highest complementarity scores \modi{and substitutability scores} with the query products.}
	\label{tab:rank}
	\tabulinesep = 1.2mm
	\begin{tabu}{cllll}
		\hline \hline
		\multicolumn{1}{l}{Query product}                                                             &   
		\multicolumn{2}{l}{Complement}        & \multicolumn{2}{l}{Substitute}      \\
		\hline
		\multirow{3}{*}{\begin{tabular}[c]{@{}c@{}}Organic \\ Blueberries\end{tabular}}     
		& 0.14 & Organic Raspberries      & 0.50  & Blueberries    \\
		& 0.059 & Organic Strawberries    & 0.13  & Green Seedless Grapes    \\
		& 0.048 & Organic Cherry Tomatoes & 0.070 & Tomatoes on the Vine \\
		\hline
		\multirow{3}{*}{\begin{tabular}[c]{@{}c@{}}Loose\\ Cucumbers\end{tabular}}            
		& 0.098 & Salad Tomatoes          & 0.54  & Organic Loose Cucumbers   \\
		& 0.089 & Baby Plum Tomatoes      & 0.21  & Courgette Spaghetti \\
		& 0.079 & Tomatoes on the Vine    & 0.18  & Sliced Runner Beans  \\
		\hline
		\multirow{3}{*}{\begin{tabular}[c]{@{}c@{}}Salad\\ Tomatoes\end{tabular}}            
		& 0.098  & Loose Cucumbers         & 0.83 & Tomatoes on the Vine \\
		& 0.063 & Iceberg Lettuce          & 0.79 & Baby Plum Tomatoes \\
		& 0.046 & Mixed Peppers            & 0.74 & Cherry Tomatoes\\
		\hline\hline
	\end{tabu}
\end{table}

\begin{figure}
    \centering
    \includegraphics[width=.9\textwidth]{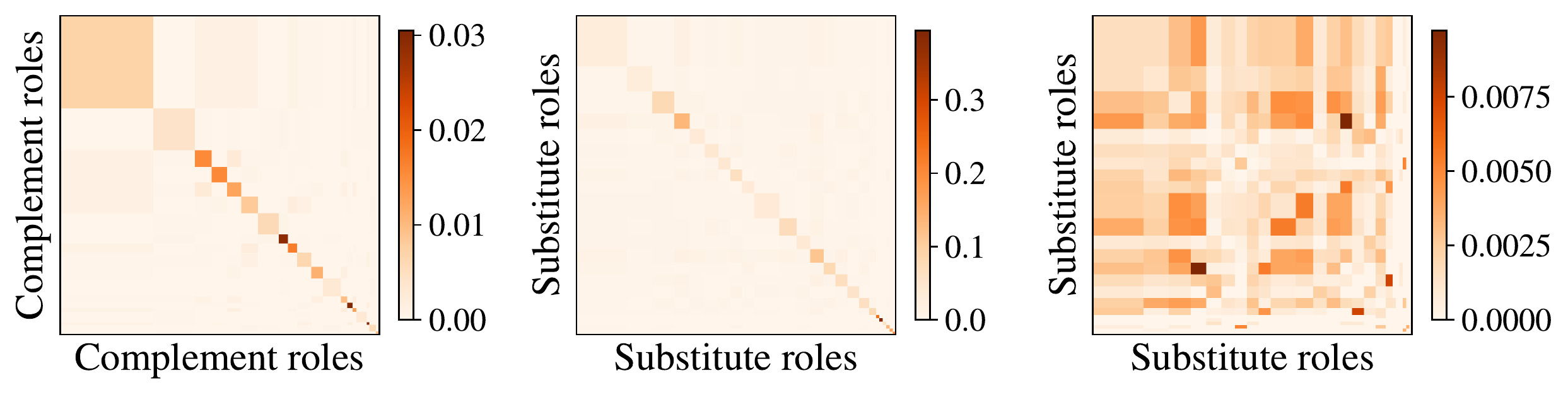}
    \caption{Role adjacencies $\mathbf{B}$, of the complement roles on the complement unipartite network $\mathbf{W}^{(c)}$ (\textbf{left}), \modi{and} of the substitute roles on the substitute unipartite network $\mathbf{W}^{(s)}$ (\textbf{middle}) and on \modi{the complement unipartite network} $\mathbf{W}^{(c)}$ (\textbf{right}), where isolated product nodes have been removed.}
    \label{fig:com_sub_wadj}
\end{figure}

We proceed for the mesoscale structure, i.e. the complement roles and substitute roles. From an averaged perspective, the complement roles and the substitute roles constitute assortative communities in the unipartite networks $\mathbf{W}^{(c)}$ and $\mathbf{W}^{(s)}$, respectively; substitute roles form disassortative communities in $\mathbf{W}^{(c)}$; see Fig. \hspace{-.2em}\ref{fig:com_sub_wadj}. The latter observation also justifies \modi{the} assumption 3. Furthermore, the overlap between the two roles is not negligible, with the normalised mutual information (NMI \cite{vinh_information_nodate}) $0.49$. Hence, as mentioned in Sect. \hspace{-.2em}\ref{sec:illustrative_example}, substitutes may appear in the same complement role by their strong complements, and complements may be assigned to the same substitute role for their strong substitutes. 

\begin{figure}
    \centering
    \begin{tabular}{cc}
        \includegraphics[width=.46\textwidth]{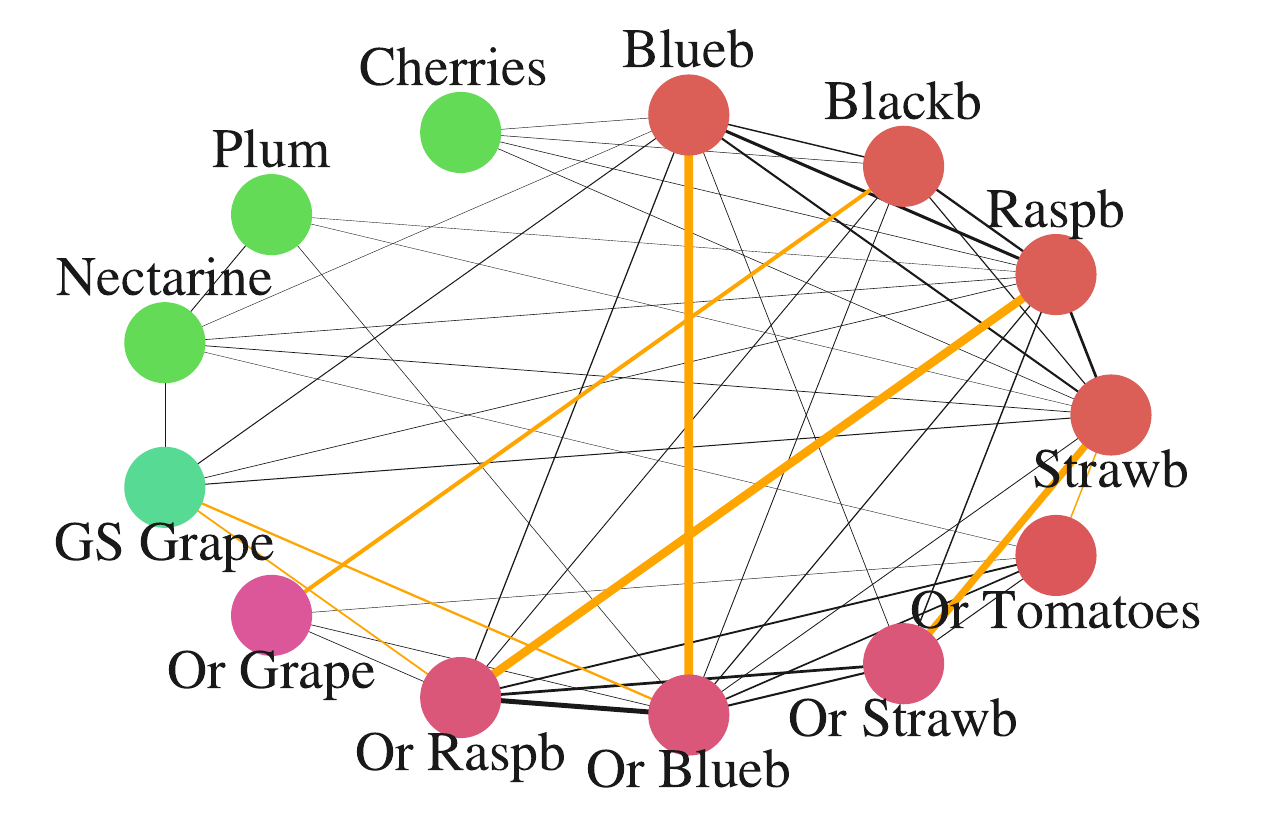} &
        \includegraphics[width=.46\textwidth]{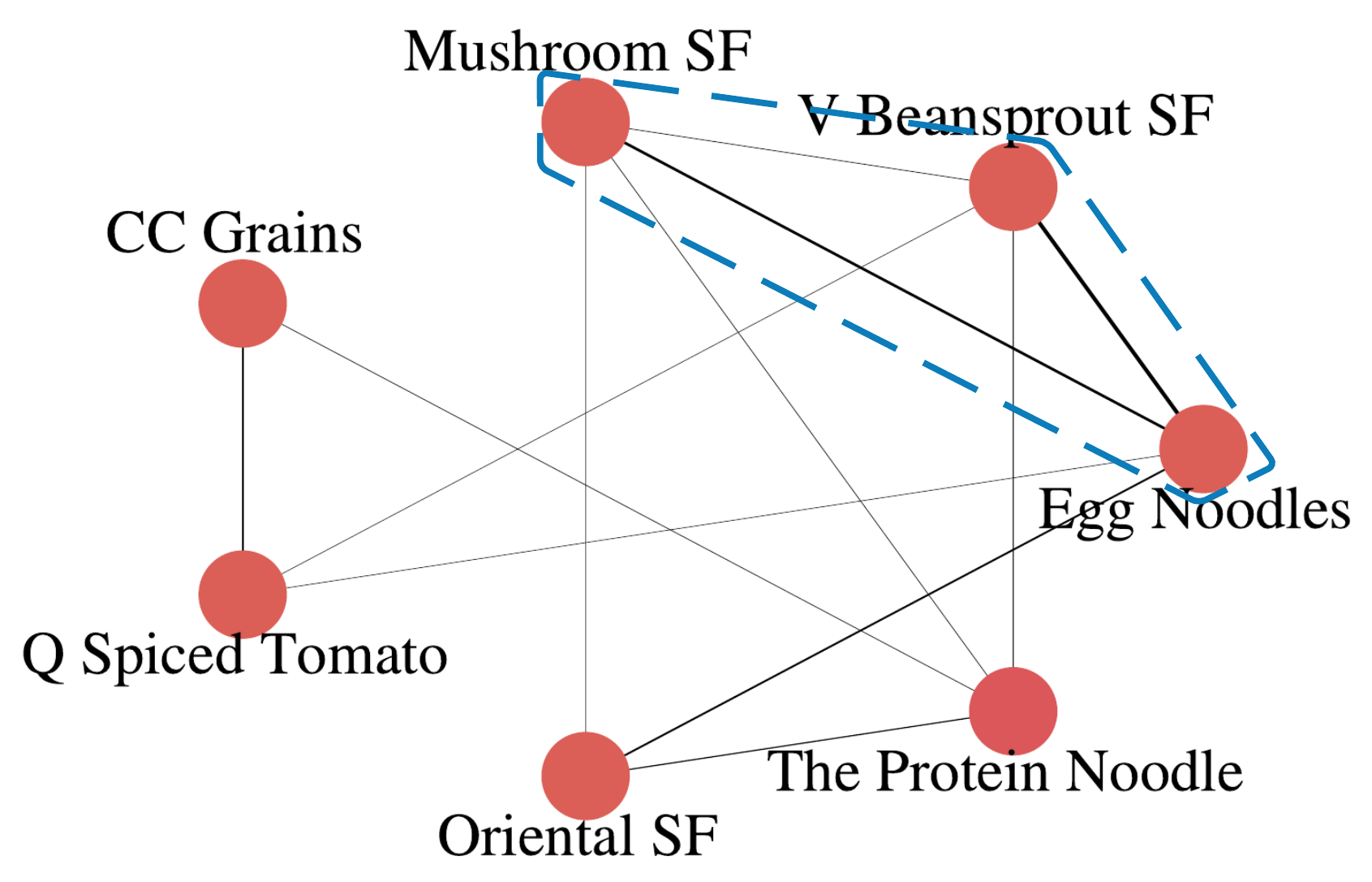}
    \end{tabular}
    \caption{Internal structure of the complement roles of berries (3) (\textbf{left}) and of stir-fry (9) (\textbf{right}), with product nodes connected by both the complementarity scores $(W_{ij}^{(c)})$ (in black) and the substitutability scores $(W_{ij}^{(s)})$ (in orange), where the line thickness corresponds to how high the scores are, and products in the same \texttt{L1} category are shown in the same colour.}
    \label{fig:com_9-11}
\end{figure}

Finally, we explore the internal structure of complement roles and substitute roles. Generally, strong complements\footnote{\footnotesize{Note we determine two products $i,j$ as complements if $W_{ij}^{(c)} > 0$, and measure their degree of complementarity (from weak to strong) by the value of $W_{ij}^{(c)}$; substitutes are treated similarly.}} do not tend to form complete graphs in the complement unipartite network $\mathbf{W}^{(c)}$, where there are many products that are complements of the same products but are not complements of each other. For example, blueberries (Blueb) and organic blueberries (Or Blueb) in the complement role of berries (3) are substitutes, but both are complements of raspberries (Raspb), stawberries (Strawb), etc; see Fig. \hspace{-.4em}\ref{fig:com_9-11}. There are also cases in which they constitute some complete graph, and further exploration indicates that these products are highly likely to be consumed together. For example, mushroom stir fry (Mushroom SF), vegetable and beansprout stir fry (V Beansprout SF), and egg noodles form a triangle in the complement role of stir-fry (9); see the blue polygon in Fig. \hspace{-.2em}\ref{fig:com_9-11}.

\begin{figure}
    \centering
    \begin{tabular}{cc}
        \includegraphics[width=.46\textwidth]{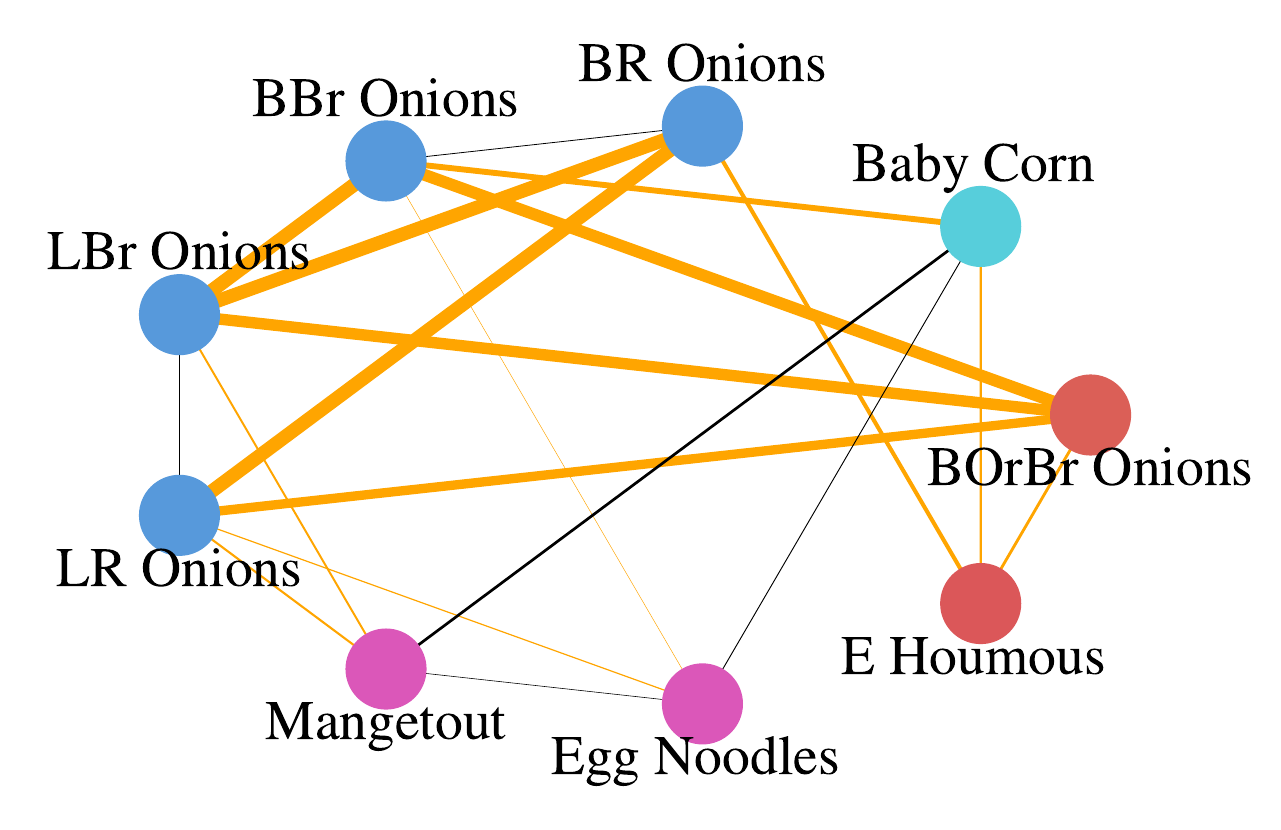} &
        \includegraphics[width=.46\textwidth]{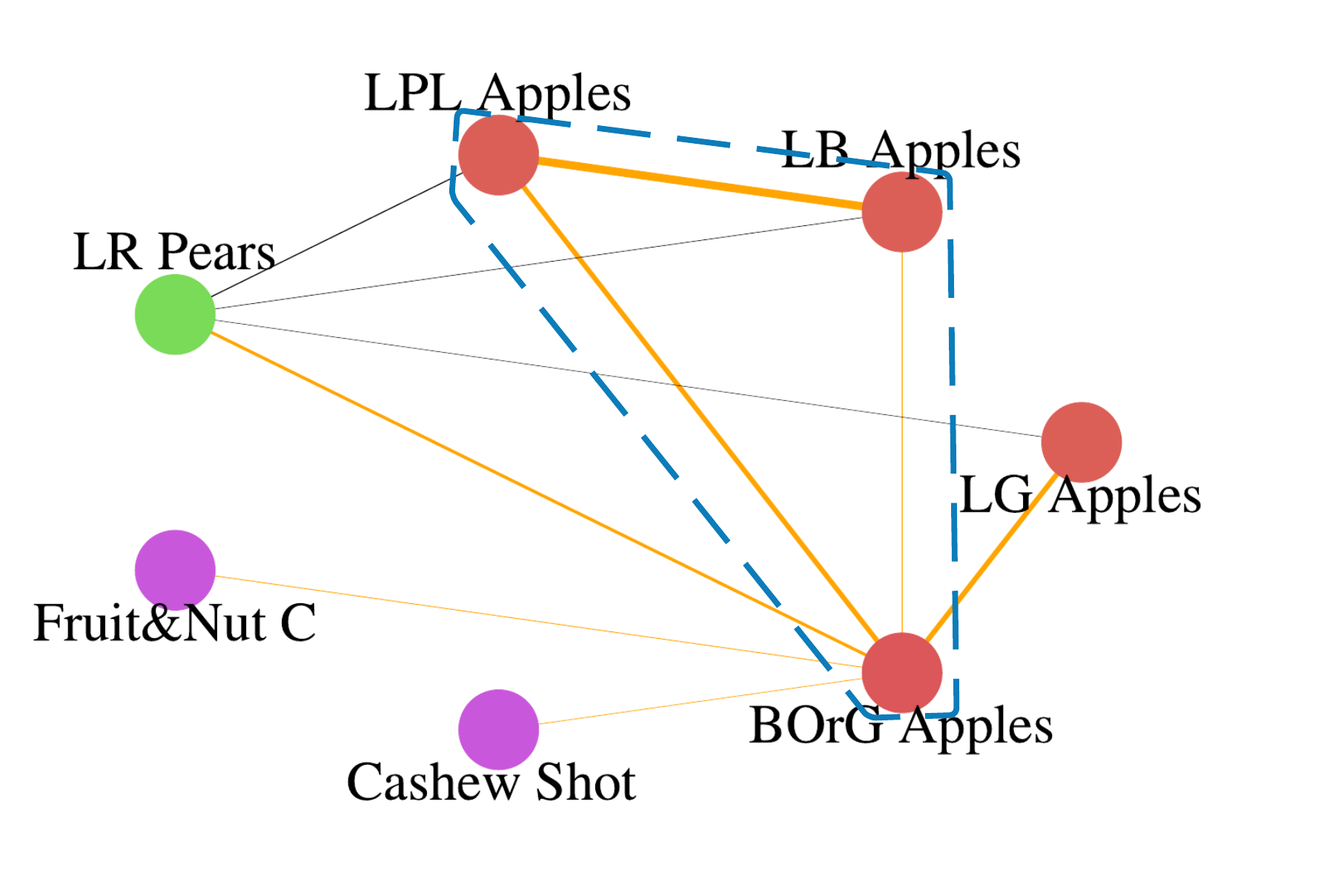}
    \end{tabular}
    \caption{Internal structure of the substitute roles of onions (4) (\textbf{left}) and of apples (23) (\textbf{right}), with product nodes connected by both the complementarity scores $(W_{ij}^{(c)})$ (in black) and the substitutability scores $(W_{ij}^{(s)})$ (in orange), where the line thickness corresponds to how high the scores are, and products in the same \texttt{L1} category are shown in the same colour.}
    \label{fig:sub_4-23}
\end{figure}

Strong substitutes are expected to form complete graphs in the substitute unipartite network $\mathbf{W}^{(s)}$, and our results are largely consistent with the expectation. For example, loose Braeburn apples (LB Apples), loose Pink Lady apples (LPL Apples), and bagged organic Gala apples (BOrG Apples) constitute a triangle in the substitute role of apples (23); see the blue polygon in Fig. \hspace{-.4em}\ref{fig:sub_4-23}. Note this expectation is only valid if the substitutes are consumed for the same purpose; if this assumption is violated, seemingly substitute products may end up being complements. For example, loose brown onions (LBr Onions) and loose red onions (LR Onions) in the substitute role of onions (4) are both substitutes of products such as bagged red onions (BR Onions) and bagged organic brown onions (BOrBr Onions), but are complements of each other; see Fig. \hspace{-.4em}\ref{fig:sub_4-23}. The difference between their quantities and their common substitutes may be the key factor here. Likewise, even with the common substitute bagged organic Gala apples (BOrG Apples), loose ripe pears (LR Pears) is a complement of loose Pink Lady apples (LPL Apples), loose Braeburn apples (LB Apples) and loose Gala apples (LG Apples). The above observations confirms the complexity of the interaction between complements and substitutes.

\subsection{Validation}
\subsubsection{Product hierarchy}
\begin{figure}
    \centering
    \includegraphics[width=.9\textwidth]{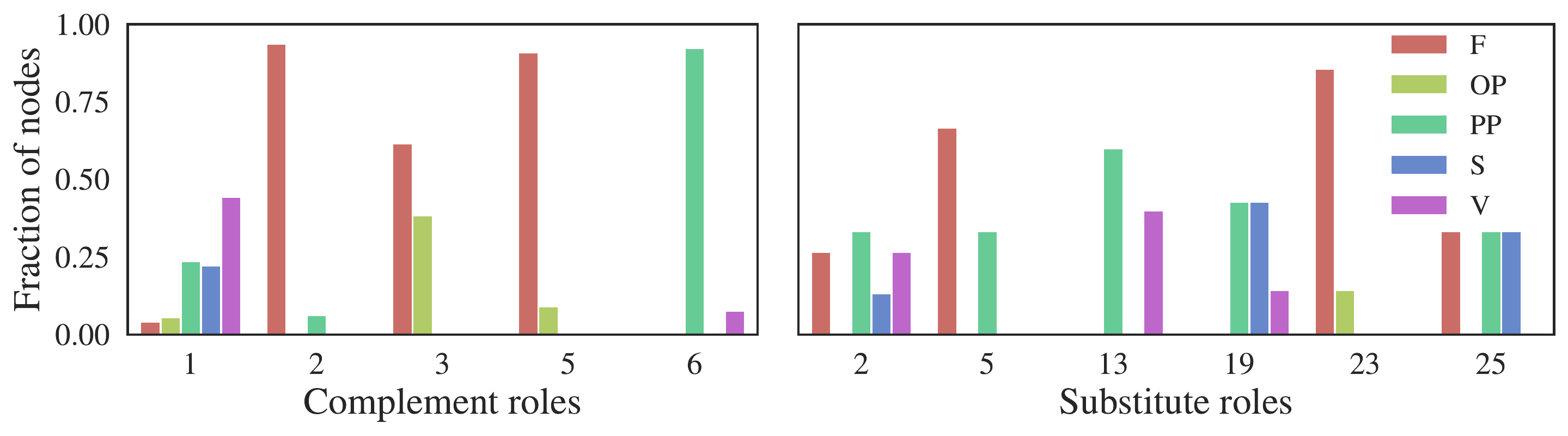}
    \caption{Proportion of the products in typical complement roles (\textbf{left}) and typical substitute roles (\textbf{right}) that fall in each \texttt{L3} category, fruit (F), organic produce (OP), prepared produce (PP), salad (S) and vegetable (V).}
    \label{fig:validate_meta_x-1}
\end{figure}

The distribution of \texttt{L3} categories in each complement role is consistent with products being complements; see Fig. \hspace{-.4em}\ref{fig:validate_meta_x-1}. Most complement roles involve more than one category, which could be explained by the complementarity across categories. For example, the complement role of nuts \& fruits (2) contains both fruit and prepared produce, the complement roles of berries \& grapes (3) and of grapes \& oranges (5) consist of both fruit and organic produce, and the complement role of potatoes, beans \& carrots (6) includes both prepared produce and vegetables. There are also complement roles only involving one category, and the related products are either in fruit or in the prepared produce category. Further, this is in agreement with the notion that products in prepared produce, for instance prepared vegetables and vegetable dips, go well together; similar for products in fruit.

The proportion of \texttt{L3} categories in each substitute role also accords with products being substitutes; see Fig. \hspace{-.2em}\ref{fig:validate_meta_x-1}. Some of them only or mostly involve prepared produce, and some others largely consist of fruit, such as the substitute role of apples (23). This agrees with the tendency of grouping products into categories based on shared characteristics. Other substitute roles contain more than one category, with one of them being prepared produce. For example, the substitute role of grapes (5) includes both fruit and prepared produce, the substitute role of carrots (13) comprises both prepared produce and vegetables, the substitute role of peppers (19) involves prepared produce, salad\footnote{\footnotesize{Here the salad category contains products like cucumber, pepper, lettuce and tomatoes.}} and vegetables, and the substitute role of avocado salad (25) is composed of fruit, prepared produce and salad. Further investigation shows that products in prepared produce include fresh-cut fruits, prepared salads and prepared vegetables, i.e. prepared versions of products in fruit, salad and vegetable categories. 

\subsubsection{Flavour compounds and recipes}
\begin{figure}
    \centering
    \includegraphics[width=.8\textwidth]{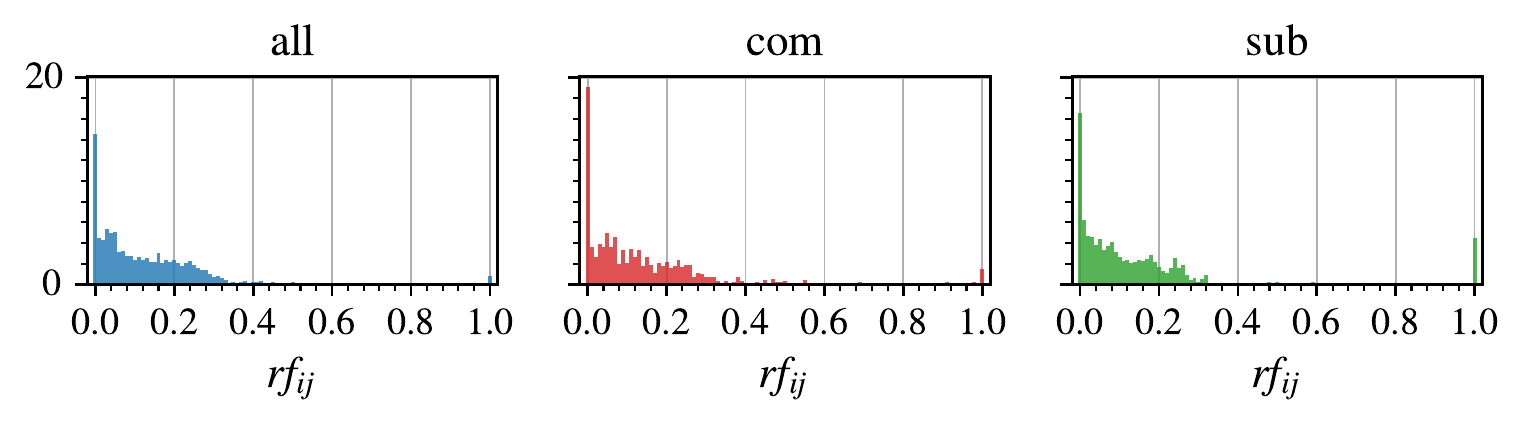}
    \caption{Distributions of the relative number of shared flavour compounds of all product pairs $\{rf_{ij}\}$ ("all"), of complementary pairs $\{rf_{ij}: W_{ij}^{(c)} > 0\}$ ("com") and of substitute pairs $\{rf_{ij}: W_{ij}^{(s)} > 0\}$ ("sub"), \modi{where probabilities of $rf_{ij}=0$ and $rf_{ij}=1$ are of interest.}}
    \label{fig:comp_3dist}
\end{figure}

We observe that the substitute pairs have a significantly higher probability to share all their flavour compounds with each other, i.e, $rf_{ij} = 1$, than all product pairs, while complementary pairs have a significantly higher probability to share no flavour compounds with each other, \modi{i.e, $rf_{ij} = 0$}; see Fig. \hspace{-.2em}\ref{fig:comp_3dist}. These characteristics are consistent with the functional definition of complements and substitutes: complements are consumed together, thus tend to have different flavours in order to accompany each other; while substitutes can replace each other, thus tend to have the same flavours. The distributions of $rf_{ij}$ values between $0$ and $1$ have roughly the same shapes for the three types of pairs. 

Further, we investigate the correlations between the relative number of shared flavour compounds $(rf_{ij})$ and \modi{the score values $(W_{ij}^{(c)}), (W_{ij}^{(s)})$; see Table \ref{tab:com_recp_corr}}. The Pearson correlation indicates that the product pairs of higher substitutability scores have a significant tendency to share larger portions of their flavour compounds, while the patterns when changing the complementarity scores is more heterogeneous, \modi{with} a mild negative correlation between the ranking of the complementarity scores and that of the relative number of shared flavour compounds.

\begin{figure}
    \centering
    \includegraphics[width=.8\textwidth]{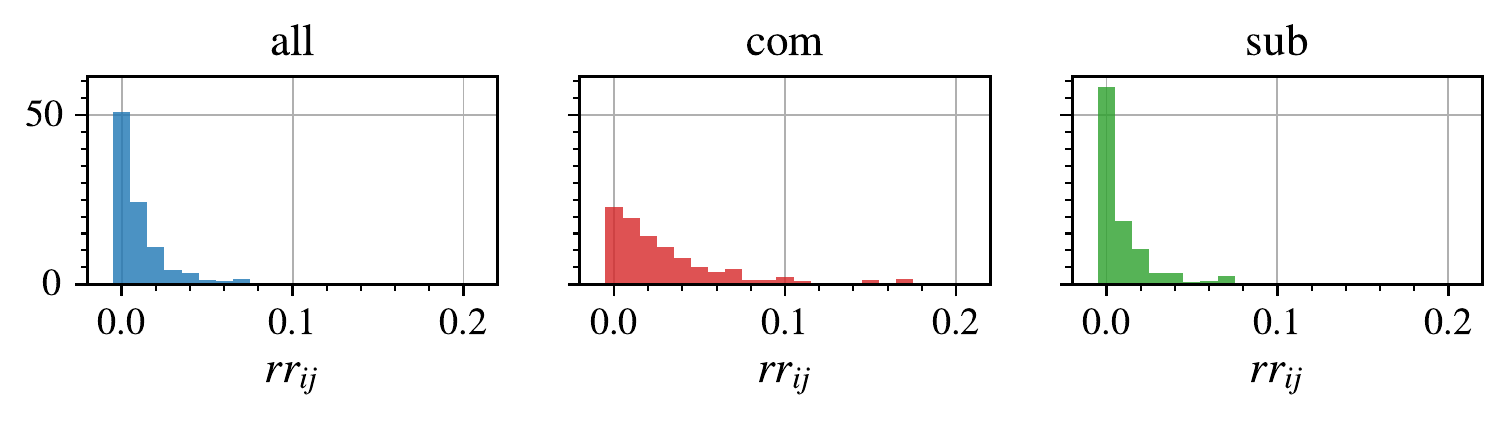}
    \caption{Distributions of the relative number of shared recipes of all product pairs $\{rr_{ij}\}$ ("all"), of complementary pairs $\{rr_{ij}: W_{ij}^{(c)} > 0\}$ ("com") and of substitute pairs $\{rr_{ij}: W_{ij}^{(s)} > 0\}$ ("sub"), \modi{where the range is chosen for visualisation purpose, while complementary pairs also have positive probabilities at values greater than $0.2$.}}
    \label{fig:recipe_3dist_single}
\end{figure}

We then discern that the complementary pairs have higher probability to co-appear in relatively more recipes, $\{rr_{ij}: W_{ij}^{(c)} > 0\}$, than all product pairs, $\{rr_{ij}\}$, while the substitute pairs have \modi{lower} probability to co-appear in relatively \modi{more} recipes, $\{rr_{ij}: W_{ij}^{(s)} > 0\}$; see Fig. \hspace{-.4em}\ref{fig:recipe_3dist_single}. Both trends are significant by the MWW test (p-values: $2.8\times 10^{-123}$ and $3.1\times 10^{-4}$ for the complementary pairs and the substitute pairs respectively). These features also accord with the interpretation of complements or substitutes from the cooking perspective: complements go \modi{well} with one another, thus are more likely to be appear in the same recipe together; while substitutes can be used in place of each other, thus tend to be cooked together with some others but not each other.

Moreover, we examine the correlations between the relative number of shared recipes $(rr_{ij})$ and \modi{the score values; see Table \ref{tab:com_recp_corr}}. The Spearman correlation suggests that product pairs of higher rankings in the complementarity scores tend to co-appear in relatively more recipes, which agrees with the Pearson correlation. The trend when increasing the substitutability ranking of product pairs is a mild propensity towards co-appearing in relatively less recipes.

Additionally, we explore the correlations between our complementarity scores $(W_{ij}^{(c)})$ and the recipe complementarity scores $(W_{ij}^{(cr)})$, and between our substitutability scores $(W_{ij}^{(s)})$ and the recipe substitutability scores $(W_{ij}^{(sr)})$; \modi{see Table \ref{tab:com_recp_corr}}. The Spearman correlations \modi{of both score pairs indicate significant positive relationships within each pair, which is consistent with the information suggested by their Pearson correlations}.
\begin{table}
    \centering
    \caption{\modi{The correlations between the scores from sales data, $(W^{(c)}_{ij}), (W^{(s)}_{ij})$, and the measures from the flavour compound and recipe data, $(rf_{ij}), (rr_{ij})$ (where the superscripts $(c)$ and $(s)$ below denote the values restricted to $\{(i,j): W^{(c)}_{ij} > 0\}$ and $\{(i,j): W^{(s)}_{ij} > 0\}$, respectively), $(W^{(cr)}_{ij}), (W^{(sr)}_{ij})$.}}
    \vspace{.5em}
    \modi{
    \begin{tabular}{c|llll}
        Score pair & Pearson &(p-value) & Spearman & (p-value)\\
        \hline
        $(W^{(c)}_{ij}) - (rf^{(c)}_{ij})$ & $0.085$ & $(4.8\times 10^{-4})$  & $-0.019$ & $(4.3\times 10^{-1})$\\
        $(W^{(s)}_{ij}) - (rf^{(s)}_{ij})$ & $\mathbf{0.50}$ & $(2.4\times 10^{-79})$  & $0.030$ & $(3.1\times 10^{-1})$ \\
        $(W^{(c)}_{ij}) - (rr^{(c)}_{ij})$ & $\mathbf{0.16}$ & $(7.2\times 10^{-8})$  & $\mathbf{0.24}$ & $(1.1 \times 10^{-16})$ \\
        $(W^{(s)}_{ij}) - (rr^{(s)}_{ij})$ & $-0.040$ & $(2.7\times 10^{-1})$ & $-0.062$ & $(8.8\times 10^{-2})$\\
        $(W^{(c)}_{ij}) - (W^{(cr)}_{ij})$ & $\mathbf{0.16}$ & $(3.8\times 10^{-8})$ & $\mathbf{0.23}$ & $(2.5\times 10^{-15})$ \\
        $(W^{(s)}_{ij}) - (W^{(sr)}_{ij})$ & $\mathbf{0.60}$ & $(4.5\times 10^{-76})$ & $\mathbf{0.23}$ & $(2.5\times 10^{-15})$ \\
    \end{tabular}}
    \label{tab:com_recp_corr}
\end{table}

Finally, we compare the complement and substitute role assignments from different data sources, in particular the sales and recipe data, \modi{where we use the NMI and adjusted mutual information (AMI \cite{vinh_information_nodate}) to measure the consistency between role assignments}; see Table \ref{tab:recipe_partition_nmi_single}. Our substitute roles (from the sales data) are more similar to the complete substitution, i.e. $l_0$ substitute roles obtained from the recipe data. Although our complement roles are more in agreement with the $l_0$ substitute roles than the $l$ complement roles by NMI, this may be caused by the number of $l_0$ substitute roles being larger than that of $l$ complement roles, since the AMI shows a significant opposite direction. \modi{To conclude,} the relatively large NMI and AMI values demonstrate the consistency between the extracted product relationship from these two different sources, and also provide evidence that customers buy products corresponding to ingredients in particular recipes. 
\begin{table}
\centering 
\caption{NMI and AMI between the partitions by the roles from the sales data (columns) and those from the recipe data (rows), where "com" and "sub" correspond to the complement roles and the substitute roles, respectively.}
\label{tab:recipe_partition_nmi_single}
\vspace{.5em}
\begin{tabular}{l|lll}
    NMI/AMI & $l_0$ sub & $l_1$ sub & $l$ com \\
    \hline 
com & $0.54$/$0.16$  &      /            & $0.36$/$0.28$    \\
sub & $0.71$/$0.21$  & $0.54$/$0.18$  &         / 
\end{tabular}
\end{table}

\section{Discussion}
\label{sec:discussion}
Extracting complements and substitutes is part of the broad family of unsupervised learning problems, since the relationship between any pair of products is unknown \cite{hastie_unsupervised_2009} (see \modi{Appendix \ref{sec:supp_unsupervised_learning}} for the detailed formulation). This makes the validation process ill-defined, as there is no ground truth. Hence in our study, not only do we compare the results with heuristic arguments based on common understanding of the product relationships, but we also resort to external data sources -- the product hierarchy data, the flavour compound and recipe data. Since these datasets focus on different aspects of the products, this is a well-grounded validation process. The seemingly heterogeneous observations from such datasets are well-explained by the product relationships, and thus provide further validation of our results.

Our assumption that complements are products purchased together significantly more frequently could appear simplistic, because it does not explicitly exclude other factors that may result in co-purchases, e.g. \hspace{-.2em}correlated preference. However, from a network perspective, these effects are expected to be removed implicitly by the statistical tests associated \modi{with} our null models. Moreover, we also propose a family of randomised measures to explicitly remove various sorts of noise \modi{effects}. Compared with the state-of-the-art, another advantage offered by a network perspective is the definition of exact criteria to determine whether products are complements or substitutes. \modi{In this article}, we have shown that both relationships can be effectively extracted from the simple notions of whether two products are purchased together significantly more frequently, or less frequently but share common strong complements (assumptions 1 and 3 in Sect. \hspace{-.2em}\ref{sec:assumptions}).

Once unipartite networks of products have been built, we may proceed from pairwise relationships to the mesoscale structure, via the notion of complement roles and substitute roles. The observations justify our assumption \modi{3} that substitutes share common strong complements. They also indicate that complement products do not generally constitute complete graphs, while substitute products typically do, though such complete graphs can be destroyed, \modi{for example,} by substitutes consumed for different purposes. These results demonstrate the possibility of the complement relationship to go beyond pairwise relationship, and also the complex interaction between complements and substitutes.  

Finally, let us emphasise that we only use basket data to extract the product relationships, without additional information such as the customer profile and the price change, information that are typically required for existing methods and may cause privacy issues \cite{de2015unique}. Our method to extract complements and substitutes is then solely based on sales data, as stated in the assumptions in Sect. \hspace{-.4em}\ref{sec:assumptions}. Hence, the quality of our results is dependent on the mutual information between the sales data (through our assumptions) and the criteria, where some discrepancy may exist. For example, there may be products that are not generally recognised as complements, but are purchased together significantly more frequently, so are treated as "complements" from the sales angle. However, most applications of product relationships are from a sales perspective, such as stocking shelves and marketing in sales promotions, and our validation further confirms the rationality of our extracted complements and substitutes.

For these reasons, we believe that the network-based approach is a promising research avenue within the field of retail. Among the research directions that this article has opened, \modi{an important one would be to consider the bipartite network from a temporal perspective, in order to explore further the connection between structure and cross-elasticity (see Appendix \ref{sec:supp_robustness})}. It would also be interesting to design a method that directly uncovers the degrees of complementarity and substitutability from the bipartite network, without any intermediary steps as it is done here, \modi{and to explore more of the directed scores, since our focus is on the symmetric ones here. Another is to characterise the products by their centrality in the projected networks, for instance by the average complementarity and substitutability scores of their relations. Moreover,} our current analysis focused on fresh food where prices changed frequently throughout the period. \modi{Yet}, we did not explicitly include price \modi{as a} factor, but either ignored its \modi{bias} or removed it by some random models \cite{athey_comtheory_1998}. In order to analyse a more general range of products in the future, it \modi{would be} necessary to incorporate price information in our framework in a meaningful way.

\section*{Declaration}
\subsection*{Acknowledgements}
  We thank Yong-Yeol Ahn for providing the flavour compound and recipe data, and for his assistance.

\subsection*{Funding}
  YT is supported by the EPSRC Centre for Doctoral Training in Industrially Focused Mathematical Modelling (EP/L015803/1) in collaboration with Tesco PLC. 

\subsection*{Abbreviations}
AMI, adjusted mutual information; BiCM, bipartite configuration model; ER, Erd\H{o}s-R\'enyi; ML, machine learning; NMI, normalised mutual information; MWW, Mann-Whitney-Wilcoxon.   

\subsection*{Availability of data and materials}
  The flavour compound and recipe data is available from \cite{Ahn_flavor_2011, yy_flavour_2011}. The other datasets generated and analysed during the current study are not publicly available due to the terms of use in Tesco PLC. 

\subsection*{Competing interests}
  The authors declare that they have no competing interests.

\subsection*{Author's contributions}
  All authors read and approved the final manuscript. 

\vspace{2em}
\bibliographystyle{ieeetr} 
\bibliography{bmc_article}      

\vspace{2em}
\appendix
\section{Bipartite Configuration Models (BiCMs)}
\label{sec:app_bicm}
\modi{The configuration model creates a network with a given degree sequence $\{d_i\}$, by assigning $d_i$ half-edges (or stubs) to each node $i$ and joining two chosen stubs uniformly at random until no more stubs are left \cite{newman_networks_2018, newman_randomg_2001}. The BiCM takes the bipartite features into account, where two degree sequences are given, dividing the nodes into two subsets, and edges are only allowed between the two subsets of nodes.} We denote the two given degree sequences $\{d_l^{(0)}\}$ and $\{d_i^{(1)}\}$ for the two subsets of nodes (i.e. part 0 and part 1). Hence, $\sum_{l=1}^{n_0}d_l^{(0)} = \sum_{i=1}^{n_1}d_i^{(1)}=m$, where $m$ is the number of edges, and $n_0, n_1$ are the numbers of nodes in parts 0 and 1, respectively. 

\subsection{Probability of Edge $(l,i)$}
\label{sec:app_bicm_edge_prob}
Consider one stub of node $l$ in part 0, the probability to connect to one of the $d_i^{(1)}$ stubs of node $i$ among all $m$ stubs in part 1 is $d_i^{(1)}/m$. Since there are $d_l^{(0)}$ stubs of $l$, if we approximate the connecting process as $d_l^{(0)}$ independent Bernoulli trials with probability $d_i^{(1)}/m$\footnote{\footnotesize{Note the actual probability depends on how many stubs in part 1 are left, and also how many stubs from node $i$ remain.}}, then the probability of nodes $l$ and $i$ to be connected, i.e. edge $(l,i)$, is, 
\begin{align*}
	p_{li} = 1 - \left(1 - \frac{d_i^{(1)}}{m}\right)^{d_l^{(0)}} \to 1 - \left(1 - d_l^{(0)}\frac{d_i^{(1)}}{m}\right) = \frac{d_l^{(0)}d_i^{(1)}}{m},\text{ as }\frac{d_i^{(1)}}{m} \to 0.
\end{align*}
We will use the approximation in the following analysis since the limit is commonly true in large networks. Note that the probability of an edge between two nodes only depends on their degrees if they are in different parts (0 otherwise). 

Since BiCMs do not exclude multi-edges, it is then important to know how probable it is to obtain a multi-edge. Suppose node $l$ and $i$ are already connected, the probability of getting another edge between them is then the probability of an edge between a node of degree $d_l^{(0)}-1$ and another of degree $d_i^{(1)}-1$. Hence, the probability of obtaining at least two edges between $l$ and $i$ is $d_l^{(0)}d_i^{(1)}(d_l^{(0)}-1)(d_i^{(1)}-1)/m^2$. Suppose the processes to form multi-edges between every pairs of nodes are independent Bernoulli trials with possibly different probabilities, then the expected number of pairs with multi-edges is
\begin{align*}
    \begin{split}
        \sum_{l=1}^{n_0}\sum_{i=1}^{n_1}&\frac{d_l^{(0)}d_i^{(1)}(d_l^{(0)}-1)(d_i^{(1)}-1)}{m^2}
        = \left(\frac{<d^{(0)2}> - <d^{(0)}>}{<d^{(0)}>}\right)\left(\frac{<d^{(1)2}> - <d^{(1)}>}{<d^{(1)}>}\right),
    \end{split}
\end{align*}
where $<d^{(z)q}> = \sum_{i=1}^{n_z}d_i^{(z)q}/n_z, z\in \{0,1\}$, is the $q$-th moment of the degree sequence $\{d^{(z)}_i\}$, and $n_z$ is the number of nodes involved. Hence the number of pairs having multi-edges stays constant as long as the moments are constant and finite, and will be negligible if the network is sufficiently large.  

\subsection{Common Neighbours}
\label{sec:app_bicm_com_nei}
The common-neighbour pattern is important in characterising nodes in bipartite networks, hence we now consider the number of common neighbours between nodes $i$ and $j$ in part 1, $cn^{(1)}_{ij}$, and that between nodes $h$ and $l$ in part 0, $cn^{(0)}_{hl}$, follows naturally.

For a node $l$ in part $0$, we know the probabilities of edges $(l,j)$ and $(l,i)$, but if $l$ is already connected to $j$, the probability to also connect to $i$ will be $(d_l^{(0)}-1)d_i^{(1)}/m$. The probability of product nodes $i,j$ sharing a transaction node $l$ is then
\begin{align*}
    p^{(1)}_{ilj} = \frac{d_l^{(0)}d_j^{(1)}(d_l^{(0)}-1)d_i^{(1)}}{m^2},
\end{align*} 
Since $i$, $j$ can have any node in part 0 as their common neighbour, if we consider the whole process as $n_0$ independent Bernoulli trials with possibly different probabilities, $cn^{(1)}_{ij}$ is then a Poisson binomial random variable, with the expected value
\begin{align*}
	\mu_{ij}^{(1)} = \sum_{l=1}^{n_0}\frac{d_l^{(0)}d_j^{(1)}(d_l^{(0)}-1)d_i^{(1)}}{m^2} = \frac{d_i^{(1)}d_j^{(1)}}{m}\frac{<d^{(0)2}> - <d^{(0)}>}{<d^{(0)}>}. 
\end{align*} 
Similarly for $h$, $l$ in part $0$, $cn^{(0)}_{hl}$ is a Poisson binomial random variable with the mean value 
\begin{align*}
	\mu_{hl}^{(0)} = \sum_{i=1}^{n_1}\frac{d_h^{(0)}d_i^{(1)}d_l^{(0)}(d_i^{(1)}-1)}{m^2} = \frac{d_h^{(0)}d_l^{(0)}}{m}\frac{<d^{(1)2}> - <d^{(1)}>}{<d^{(1)}>}.
\end{align*}
Hence the expected number of common neighbours is dependent on both the degrees of the two nodes, and the moments of the other part's degree sequence.

\section{Poisson binomial distribution}
\label{sec:app_poisson_binomial}
The Poisson binomial distribution is the discrete probability distribution of a sum of independent Bernoulli random variables that are not necessarily identically distributed, i.e. \hspace{-.4em}with parameters $p_1, p_2, ..., p_n$ that are possibly different. 

Since a Poisson binomial random variable is the sum of $n$ independent Bernoulli distributed variables, its mean and variance are simply the sums of the means and the variances of corresponding Bernoulli distributions, respectively, i.e. $\mu = \sum_{i=1}^{n}p_i,\ \sigma^2 = \sum_{i=1}^{n}p_i(1-p_i)$. 

\subsection{Poisson Approximation}
\label{sec:app_poisson_approxi}
The distribution of a Poisson binomial random variable can be approximated by that of a Poisson random variable, $Pois(\lambda)$, with the same mean value $\lambda = \mu$.
\begin{theorem}[Le Cam's Theorem]
Let $\{X_j: j=1, 2, \dots\}$ be a family of independent random variables. Assume $X_j\sim Bernoulli(p_j)$, then $Y = \sum_{j}X_j$ is a Poisson binomial random variable. Let $\lambda=\sum_{j}p_j$, $\omega = \sum_{j}p_j^2/\lambda$ and $\alpha=\sup_j p_j$. Denote by $Q$ the distribution of \hspace{.3em}$Y$ and $P$ the Poisson distribution, $Poisson(\lambda)$. There exist constants $D_1$ and $D_2$ such that 
\begin{enumerate}
    \item For all values of $p_j$, 
    $$||P - Q|| \le 2\lambda\omega,$$
    and 
    $$||P - Q|| \le D_1\alpha.$$
    \item If $4\alpha \le 1$, then 
    $$||P - Q|| \le D_2\omega.$$
\end{enumerate}
The constant $D_1$ is inferior to $9$ and the constant $D_2$ is inferior to $16$.
\label{theorem:le_cam}
\end{theorem}
See \cite{lecam_poisson_1960} for the detailed proof, but we can easily see that the variance of $Y$ approaches its mean while the Bernoulli probabilities $p_j$s approach $0$. 

In our case, the composing probability is $p_{ilj}$, the likelihood for each pair of product nodes $i$ and $j$ to both connect with transaction node $l$ in the BiCM,
\begin{align*}
    p_{ilj} = \frac{d_i^{(p)}d_l^{(t)}d_j^{(p)}(d_l^{(t)}-1)}{m^2},
\end{align*}
where $d_i^{(p)}$ is the degree of product node $i$, $d_l^{(t)}$ is the degree of transaction node $l$, and $m$ is the number of edges in the bipartite network. Hence, we compute the maximum composing probability for each pair of product nodes $i$ and $j$, $p^*_{ij} = \max_{l}p_{ilj}$, and find that most pairs satisfy the condition $p^*_{ij} \le 0.25$, with the exception of only $32$ out of all $31878$ pairs. Further investigation shows that such excepted pairs all include nodes of degree higher than $2475$ ($3$ out of all $253$ nodes), i.e. hubs, and most of their composing probabilities $p_{ilj}$s are much smaller than $0.25$\footnote{\footnotesize{\modi{$p^*_{ij}$ is obtained through the maximum degree of transaction nodes ($20$), but more than $99.8\%$ of such degrees is not larger than its half ($10$), and $97.0\%$ is inferior to its quarter ($5$).}}}.

Finally, we evaluate the error bound values. For the pairs satisfying the condition, we use the tighter bound of $D_2\omega_{ij}$ where $\omega_{ij} = \sum_l p_{ilj}^2/(\sum_h p_{ihj})$. We find that the maximum value of $\omega_{ij}$s is around $0.021$, with more than $97.2\%$ pairs of $\omega_{ij} \le 0.003$ and more than $89.3\%$ pairs of $\omega_{ij} \le 0.001$, thus the Poisson approximation is guaranteed to perform well. For those that do not satisfy the condition, we analyse the looser bounds $2\lambda_{ij}\omega_{ij}$ and $D_1p^*_{ij}$, where $\lambda_{ij} = \sum_l p_{ilj}$. We find that $\lambda_{ij}\omega_{ij}$s are all larger than $2$ (and $p^*_{ij} > 0.25$, as we know), thus the Poisson approximation could be misleading for these small number of pairs. Hence, we provide the comparison between the Poisson approximation and the Chernoff bounds, an alternative approximation method, in the following Appendix \ref{sec:supp_chernoff_bounds}, in order to show that the Poisson approximation has comparable performance in the above product pairs of worse bounds. Together with the guaranteed good performance for most pairs, these are the reasons to use the Poisson approximation in Sect. \hspace{-.2em}\ref{sec:bicm}.

\subsection{Chernoff Bounds}
\label{sec:supp_chernoff_bounds}
The probability that a Poisson binomial distributed variable $X$ gets large can be bounded by the Chernoff bound for the upper tail, where for $x \ge \mu$,
\begin{align*}
	\mathbb{P}(X \ge x)\le \exp(x - \mu - x\log(x/\mu)).
\end{align*}
\begin{proof}
	By Markov inequality, for $t \ge 0$,
	\begin{align*}
		\mathbb{P}(X \ge x) = \mathbb{P}(e^{tX} \ge e^{tx}) \le \frac{\mathbb{E}[e^{tX}]}{e^{tx}}. 
	\end{align*}
	$\mathbb{E}[e^{tX}]$ is the moment generating function of a Poisson binomial variable, i.e. $X = \sum_{i=1}^{n}Y_i$ where $Y_i$s are independent \modi{of each other, and $Y_i\sim Bernoulli(p_i)$}, thus,
	\begin{align*}
		\mathbb{E}[e^{tX}] = \prod_{i=1}^{n}\mathbb{E}[e^{tY_i}] = \prod_{i=1}^{n}(1 - p_i + p_ie^{t}). 
	\end{align*} 
	Since $1 + y \le e^y,\ \forall y$, 
	\begin{align*}
		1 - p_i + p_ie^{t} \le e^{p_i(e^t - 1)}.
	\end{align*}
	Hence, 
	\begin{align*}
	\mathbb{P}(X \ge x) 
	&\le \exp(-tx)\prod_{i=1}^{n}\exp(p_i(e^t - 1))\\
	&= \exp(-tx + \sum_{i=1}^{n}p_i(e^t - 1))\\
	&= \exp(-x\log(\frac{x}{\mu}) + x - \mu),
	\end{align*}
	where we choose $t = \log(x/(\sum_{i=1}^{n}p_i)) = \log(x/\mu)$ \modi{which minimises the above upper bound w.r.t. $t$}.
\end{proof}
\modi{Similarly, the probability that a Poisson binomial distributed variable $X$ gets small can be bounded by the Chernoff bound for the lower tail, where for $0 < x < \mu$,
\begin{align*}
    \mathbb{P}(X \le x) \le \exp(x - \mu + x\log(\mu/x)).
\end{align*}
\begin{proof}
    By Markov inequality, for $t\ge 0$, 
    \begin{align*}
    \mathbb{P}(X \le x) = \mathbb{P}(-X \ge -x) = \mathbb{P}(e^{-tX} \ge e^{-tx}) \le \frac{\mathbb{E}[e^{-tX}]}{e^{-tx}}. 
    \end{align*}
    Then the proof follows the same as the previous one, with $t = \log(\mu/x)$.
\end{proof}}

\modi{\subsubsection{Accuracy}
The focus of most literature is not on exploring the theoretical guarantee for the accuracy of the Chernoff bounds, but rather on finding better Chernoff-like exponential bounds. However, we provide the proof of the exact form we use here. The inequality comes from two sources: (i) \textit{Markov inequality}, where if $Z$ is a non-negative random variable, $\mathbb{P}(Z\ge z) \le \mathbb{E}[Z]/z,\ \forall z > 0$, and here $Z = \exp(\log(x/\mathbb{E}[X])X) = (x/\mathbb{E}[X])^X$ and $z=(x/\mathbb{E}[X])^x$; (ii) $1 + y \le \exp(y),\ \forall y$, and here $y = p_i(x/\mathbb{E}[X] - 1)$ for each composing Bernoulli probability $p_i$. Accordingly, for the Chernoff bounds to be tight or exact, we need the following two conditions: (i) $Z$ can only have positive probabilities at $0$ or $z$, i.e. $X$ only have positive probability at $x$, and the more concentrated its distribution is at the value, the tighter the Markov inequality; (ii) $y=0$, i.e. $x=\mathbb{E}[X]$, the interesting value for comparison, $x$, is equal to the mean, $\mathbb{E}[X]$, since $p_i > 0,\ \forall i$, and the closer $x$ is to $\mathbb{E}[X]$ and/or each $p_i$ is to $0$, the tighter the inequality.}

\modi{In our analysis, the Poisson binomial random variable is the number of common neighbours between each pair of nodes $i$ and $j$ (in the same part) in BiCMs in Sect. \hspace{-.2em}\ref{sec:bicm}, $X_{ij} = \sum_{l}X_{ilj}$ with $X_{ilj}\sim Bernoulli(p_{ilj})$, and the value of interest is the actual number of common neighbours, $cn_{ij}$. The purpose of using the Chernoff bounds is to test whether $cn_{ij}$ is significant. Hence, for condition (i), it would be hard for each $X_{ij}$ to have support only containing $cn_{ij}$; for condition (ii), the only possibility is from the $p_{ilj}$s being sufficiently small, since $cn_{ij}$ should be far from the expected value to be significant. Hence, the Chernoff bounds are generally loose. The reasons why we consider the Chernoff bounds here are (i) to have valid upper bounds that can be evaluated efficiently, and (ii) that the relatively large estimated value can be balanced by slightly large significance values in our analysis. However, the lack of theoretical guarantee for the accuracy of the Chernoff bounds does make it less attractive from the theoretical angle.}

\modi{\subsection{Comparison: Chernoff Bounds versus Poisson Approximation}
Here we compare the results from the sales transaction data obtained by applying our framework with the Poisson approximation and with the Chernoff bounds, in order to explore whether the Chernoff bounds provide similar results to the Poisson approximation, and also whether the Poisson approximation is comparable in the pairs of worse error bounds. Here, we choose the same reference significance level $0.05$ in both cases, focus on the original measure, as in Sect. \hspace{-.2em}\ref{sec:results_salesdata}, and keep the same reference threshold quantiles $0, 0.7$ for the degrees of complementarity and substitutability, respectively. Following our framework, a higher significance level is chosen for the Chernoff bounds, see Table \ref{tab:supp_poisson_params}.} 
\begin{table}[h]
    \centering
    \modi{
    \begin{tabular}{c|cccc}
       Method    & $\alpha_m$ & $q_c$ ($\theta_c$) & $\alpha_l$ & $q_s$\\
       \hline
       Chernoff  & $0.04$     & $0.15$ ($0.012$) & $0.5$ & $0$\\
       Poisson   & $0.01$     & $0.3$ ($0.010$) & $0.25$ & $0$\\
       ER        & $0.01$     & $0.35$ ($0.011$) & $0.2$ & $0$
    \end{tabular}}
    \caption{Parameters chosen for our methods using the BiCM with the Chernoff bounds, and the Poisson approximation, as well as the variant of ER model for reference.}
    \label{tab:supp_poisson_params}
\end{table}

\begin{figure}[h]
    \centering
    \includegraphics[height=.2\textheight]{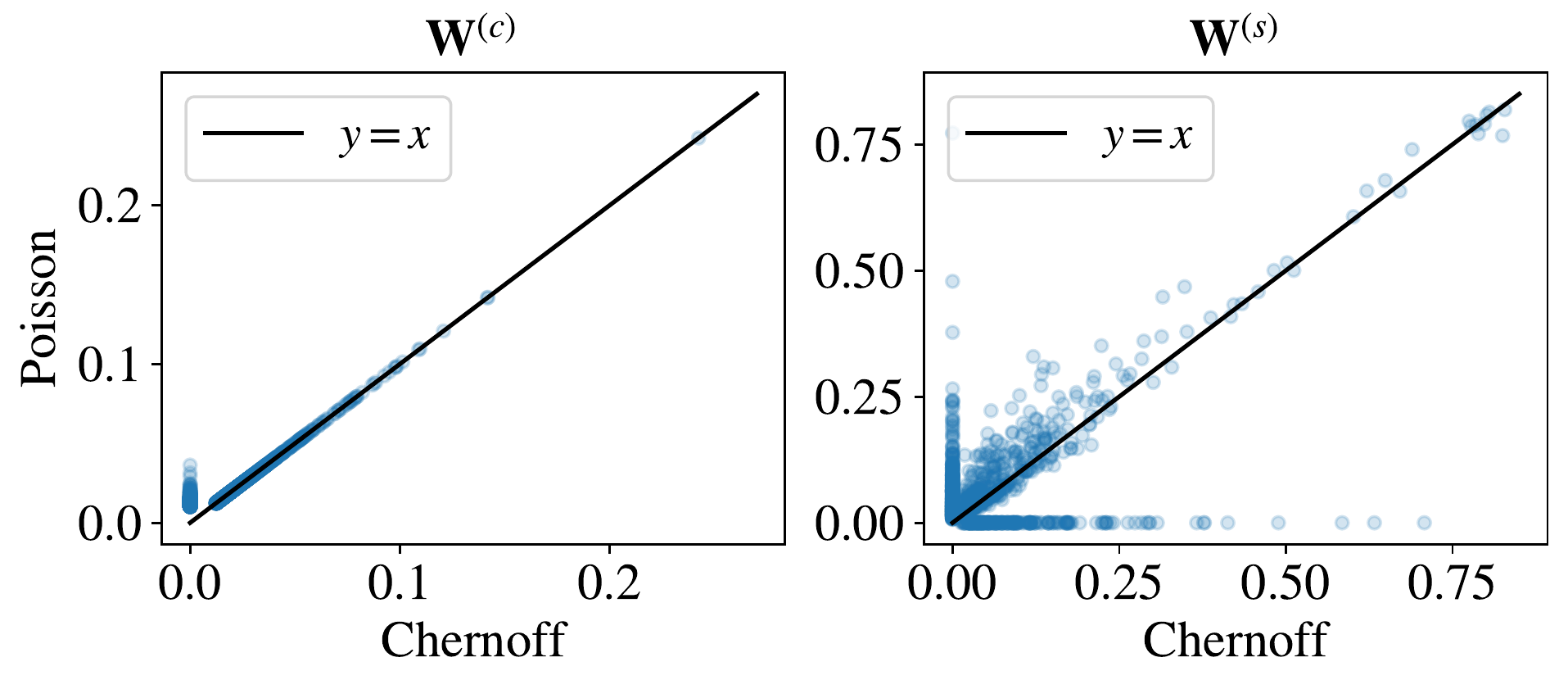}
    \caption{\modi{Scatter plots of the complementarity scores (left) and the substitutability scores (right) from the Chernoff bounds (x-axis) and the Poisson approximation (y-axis), where the identity line $y=x$ is shown in black for reference.}}
    \label{fig:supp_scatter_C-P}
\end{figure}

\modi{Comparing the score values from the two methods, they mostly agree with each other, since the scatter plots approach the identity line; see Fig. \hspace{-.2em}\ref{fig:supp_scatter_C-P} where each point is a product pair. The complementarity scores from the Poisson approximation have more nonzero values ($952$ out of all $31846$ points\footnote{\footnotesize{In the plot, we only show the $3376$ points that have nonzero values in either case.}} with $x=0, y>0$ in the left plot), where $36$ of them are caused by the discrepancy in the approximated values (rather than the difference in thresholds), and $35$ of them have $p_{ij}^* \coloneqq \max_{l}p_{ilj} \le 0.25$, which indicates that the Poisson approximation is reliable (see Appendix \ref{sec:app_poisson_approxi}). For the substitutability scores, there are $391$ points with positive score values from the Poisson approximation but zero from the Chernoff bounds (i.e. with $x=0, y>0$ in the right plot), where $67$ of them are due to the discrepancy between the approximated values (rather than carried from the difference in the complementarity scores, i.e. share no complements), and $66$ of them have $p_{ij}^*\le 0.25$; all $194$ points with positive score values from the Chernoff bounds but zero from the Poisson approximation result from the different choice of significance levels, $\alpha_l$. Hence, (i) the behaviour of the Poisson approximation largely agrees with the Chernoff bounds, but (ii) the discrepancy between them does exist, where, with a theoretical guarantee from Le Cam's theorem \cite{lecam_poisson_1960}, we can show that the Poisson approximation is expected to mostly perform well.}

\modi{The difference in scores does not cause a huge deviation in the role structure, with normalised mutual information (NMI) $0.43$ and $0.83$ and adjusted mutual information (AMI) $0.30$ and $0.36$ for complements and substitutes, respectively.}

\begin{figure}[h]
    \centering
    \includegraphics[height=.2\textheight]{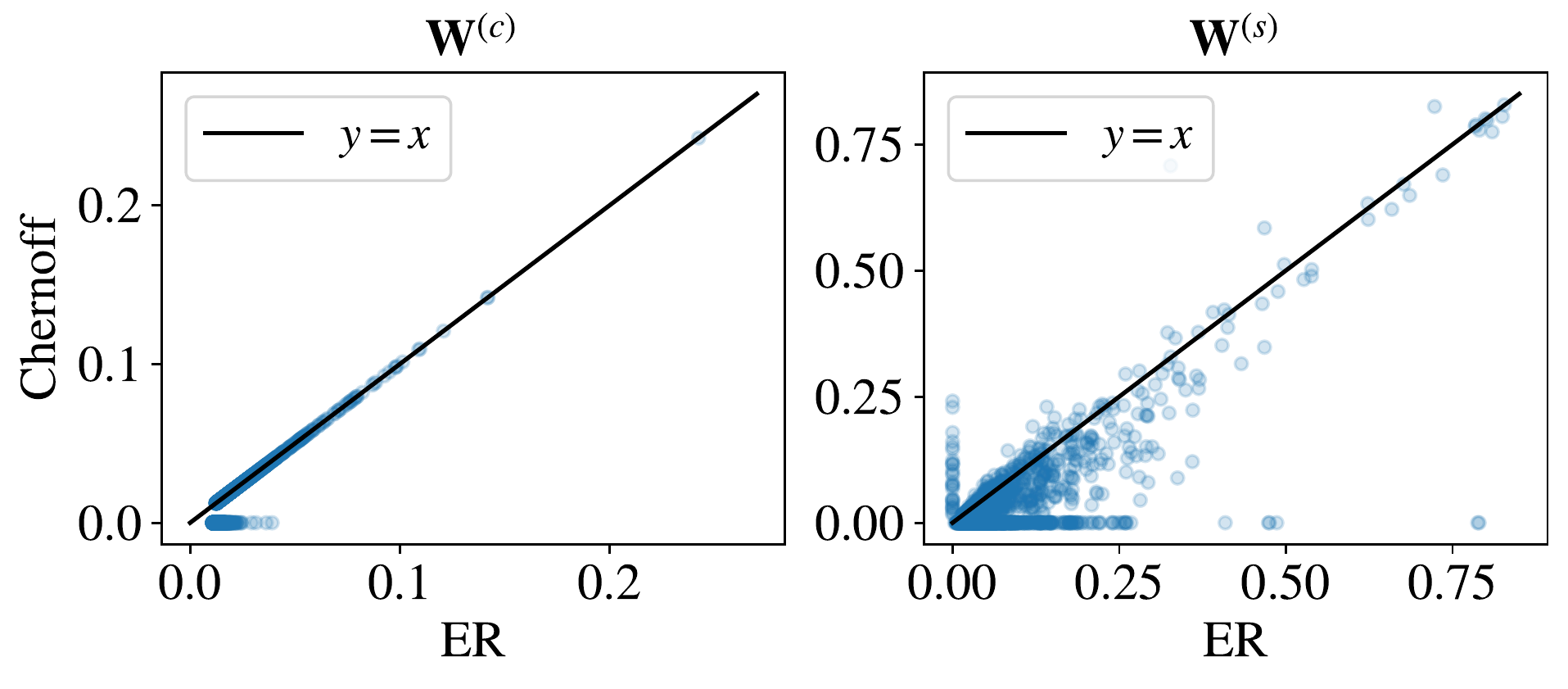}
    \caption{\modi{Scatter plots of the complementarity scores (left) and the substitutability scores (right) from the variant of ER model (x-axis) and the BiCM with the Chernoff bounds (y-axis), where the identity line $y=x$ is shown in black for reference.}}
    \label{fig:supp_scatter_E-C}
\end{figure}

\begin{figure}[h]
    \centering
    \includegraphics[height=.2\textheight]{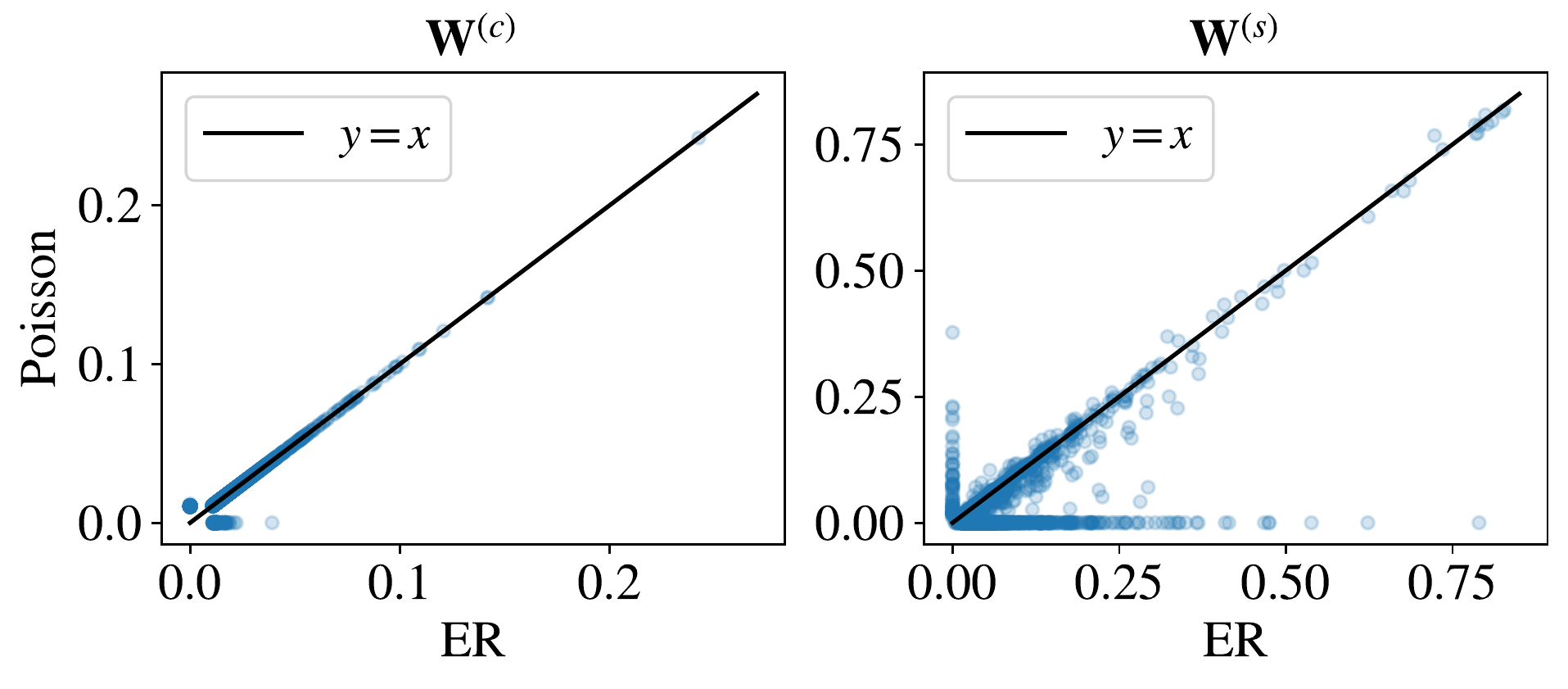}
    \caption{\modi{Scatter plots of the complementarity scores (left) and the substitutability scores (right) from the variant of ER model (x-axis) and the BiCM with the Poisson approximation (y-axis), where the identity line $y=x$ is shown in black for reference.}}
    \label{fig:supp_scatter_E-P}
\end{figure}

It is then interesting to compare both methods with the variant of ER models whose results are shown in Sect. \hspace{-.2em}\ref{sec:results_salesdata}. We can see that both are largely consistent with the results from the variant of ER model, with their scatter plots approaching the identity line; see Figs. \hspace{-.2em}\ref{fig:supp_scatter_E-C} and \ref{fig:supp_scatter_E-P}. The difference between the points of nonzero score values from the variant of ER model and the BiCM, i.e. different significant relationships, can be explained by the different underlying distributions, and these values are mostly small.

Furthermore, the resulting complement and substitute roles from the variant of ER model are not far from the results from the BiCM with both methods: those from the Poisson approximation are closer, with NMI $0.97$ and $0.79$ (AMI $0.97$ and $0.35$) respectively, than those from the Chernoff bounds, with NMI $0.43$ and $0.82$ (AMI $0.29$ and $0.41$) respectively. Hence, we only show the results from the variant of ER model in Sect. \hspace{-.2em}\ref{sec:results_salesdata}. Indeed, we can compare the expected number of common neighbours between each pair of product nodes $i$ and $j$, where in the variant of ER model, we have
\begin{align*}
    n_t\hat{p}_i\hat{p}_j = \frac{d_i^{(p)}d_j^{(p)}}{n_t},
\end{align*}
where $d_i^{(p)}$ is the degree of product node $i$ and $n_t$ is the number of transaction nodes, while in the BiCM, we have
\begin{align*}
    \sum_{l=1}^{n_t}p_{ilj} = \sum_{l=1}^{n_t}\frac{d_i^{(p)}d_l^{(t)}d_j^{(p)}(d_l^{(t)}-1)}{m^2} = \frac{d_i^{(p)}d_j^{(p)}}{m}\frac{<d^{(t)2}> - <d^{(t)}>}{<d^{(t)}>},
\end{align*}
where $d_l^{(t)}$ is the degree of transaction node $l$, $<d^{(t)}> = (\sum_{l=1}^{n_t}d_l^{(t)})/n_t$, and $<d^{(t)2}> = (\sum_{l=1}^{n_t}d_l^{(t)2})/n_t$. Then, they will be close to each other if $(<d^{(t)2}> - <d^{(t)}>)/<d^{(t)}>$ is close to $m/n_t$, and in our sales transaction data they are $1.91$ and $1.82$ respectively. Hence, the expected values are close to each other, and the results are then not expected to deviate largely from each other.

To conclude, the resulting score values from the Chernoff bounds largely agree with those from the Poisson approximation, and although there is slight discrepancy between the scores from the two methods, it does not cause substantial difference in their role structure. Meanwhile, such difference mostly exists in the pairs where the Poisson approximation is guaranteed to perform well, thus the Poisson approximation has comparable performance in the pairs of worse error bounds.

\section{Sales Data}
\label{sec:app_sales_data}
\begin{figure}[htbp]
    \centering
    \includegraphics[width=.95\textwidth]{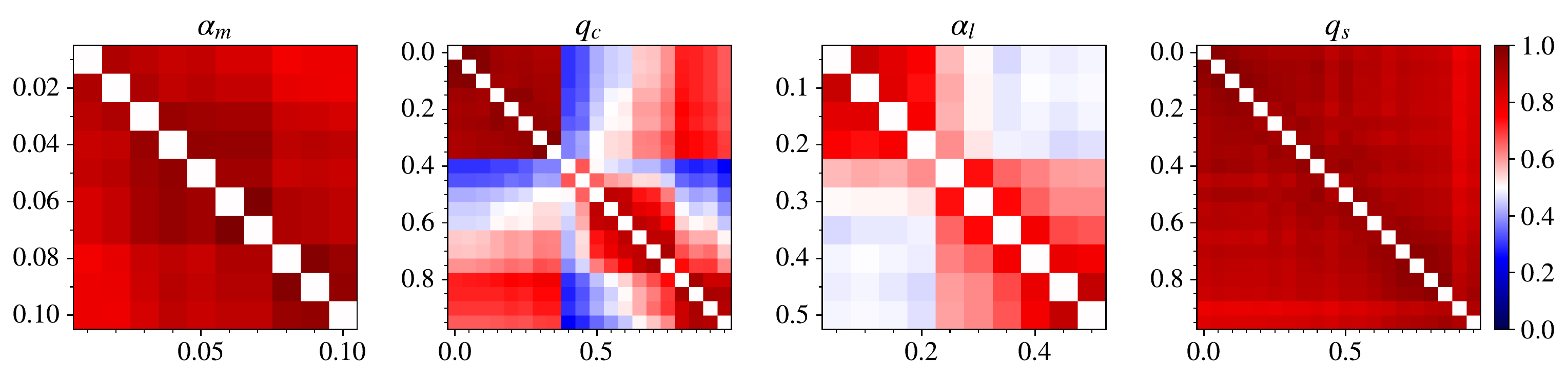}
    \caption{Pairwise NMI between the partitions of $\mathbf{W}^{(c)}$ varying $\alpha_m$ (\textbf{leftmost}) and varying the threshold quantile $q_c$ with $\alpha_m = 0.01$ (\textbf{middle-left}), and of $\mathbf{W}^{(s)}$ varying $\alpha_l$ with $\alpha_m = 0.01, q_c = 0.35$ (\textbf{middle-right}) and varying the threshold quantile $q_s$ with $\alpha_m = 0.01, q_c = 0.35, \alpha_l = 0.2$ (\textbf{right-most}), where the original measure is used and axis labels are shown in their titles.}
    \label{fig:EA_nmi_orig}
\end{figure}

We give details of how to calibrate the parameters here, including the significance levels, $\alpha_m$ and $\alpha_l$, and the thresholds, $\theta_c$ for $\mathbf{W}^{(c)}$ and $\theta_s$ for $\mathbf{W}^{(s)}$. We interpret the criterion of maintaining the same community structure as the NMI between the two underlying partitions being greater than $0.8$, and use the map equation to detect the communities. This value is chosen for sufficient consistency but limited freedom of variance between partitions \cite{Lancichinetti_comparecd_2009}. \modi{Note our} variant of bipartite ER models \modi{is used} as the underlying null model, and we choose $0.05$ as the baseline significance level in both cases. We select the thresholds $\theta_c, \theta_s$ by the quantiles of the nonzero values, $q_c, q_s$, and use $0, 0.7$ as the baseline threshold quantiles for $\mathbf{W}^{(c)}, \mathbf{W}^{(s)}$, respectively. Thus \modi{by applying the extra rules of significance-level and threshold selection}, we obtain $\alpha_m = 0.01$, $\alpha_l = 0.2$, $q_c = 0.35$ with $\theta_c = 0.011$ and $q_s = 0$, for the scores induced by the original measure; see Fig. \hspace{-.2em}\ref{fig:EA_nmi_orig}. \modi{The results for the scores induced by the randomised configuration measure are exactly the same as the above ones, (which can be inferred from the score values being very close to each other, as in Table \ref{tab:rank_rand}) thus is omitted here.} The complementarity scores $\mathbf{W}^{(c)}$, and the substitutability scores $\mathbf{W}^{(s)}$, only refer to the scores after thresholding.

Also, we show the pairwise rankings from the scores induced by the randomised configuration measure. Comparing with Table \ref{tab:rank}, we find that the results from \modi{the randomised configuration measure agree with those from the original measure. Plus, the score values induced by both measures are very close to each other, which is why they have exactly the same community structure in the parameter calibration phase, as mentioned before.} 
\begin{table}[ht]
    \centering
	\caption{Products of the three highest complementarity scores and substitutability scores with the query products, where the scores are induced by the randomised configuration measure.}
	\label{tab:rank_rand}
	\tabulinesep = 1.2mm
	\begin{tabu}{cllll}
		\hline\hline
		\multicolumn{1}{l}{Query product}                                                       & \multicolumn{2}{l}{Complement} & \multicolumn{2}{l}{Substitute} \\
		\hline
		\multirow{3}{*}{\begin{tabular}[c]{@{}c@{}}Organic \\ Blueberries\end{tabular}}       
		& 0.14  & Organic Raspberries    & 0.50  & Blueberries  \\
		& 0.058 & Organic Strawberries   & 0.13  & Green Seedless Grapes  \\
		& 0.047 & Organic Cherry Tomatoes   & 0.070 & Tomatoes on the Vine \\
		\hline
		\multirow{3}{*}{\begin{tabular}[c]{@{}c@{}}Loose\\ Cucumbers\end{tabular}}              
		& 0.095 & Salad Tomatoes  & 0.54  & Organic Loose Cucumbers \\
		& 0.085 & Baby Plum Tomatoes    & 0.21  & Courgette Spaghetti   \\
		& 0.077 & Tomatoes on the Vine    & 0.18 & Sliced Runner Beans      \\
		\hline
		\multirow{3}{*}{\begin{tabular}[c]{@{}c@{}}Salad\\ Tomatoes\end{tabular}}              
		& 0.095  & Loose Cucumbers    & 0.83 & Tomatoes on the Vine \\
		& 0.062 & Iceberg Lettuce      & 0.79 & Baby Plum Tomatoes \\
		& 0.045 & Mixed Peppers         & 0.73 & Cherry Tomatoes \\
		\hline\hline             
	\end{tabu}
\end{table}

\section{Robustness}
\label{sec:supp_robustness}
Our method is robust to temporal shifts on the condition that the underlying customers maintain their purchase habits during different time periods, i.e. they keep treating certain products as complements or substitutes. It is not necessarily true for every time period, and from our industrial collaborators, customers do change behaviours over time. Hence, it is a promising direction to further incorporate the temporal features of our scores, e.g. how the scores change over time, as mentioned in Sect. \hspace{-.2em}\ref{sec:discussion}.

\modi{For our current work, it is also interesting to explore how the scores change within the three-month period we chose. Hence, (i) we split our three-month sales transaction dataset into two one-month-and-a-half datasets; (ii) we then compare the results from the two datasets, in terms of score values and roles. Note that we now have a shorter time period, thus noise will play a relatively larger role in the analysis. Hence, we further filter our products to be sold at least once a week, which brings our analysis down to $169$ products, and $19498, 19670$ transactions in splits 1, 2 respectively. As in Sect. \hspace{-.2em}\ref{sec:results_salesdata}, we use the variant of ER model as the underlying null model, and compare the
results from the original measure, with the same choice of reference significance level ($0.05$ in both cases) and reference threshold quantiles ($0, 0.7$ for the degrees of complementarity and substituteability, respectively).}
\begin{table}[h]
    \centering
    \modi{
    \begin{tabular}{c|cccc}
       Split & $\alpha_m$ & $q_c$ ($\theta_c$) & $\alpha_l$ & $q_s$\\
       \hline
       1     & $0.01$     & $0.35$ ($0.016$) & $0.25$ & $0$\\
       2   & $0.01$       & $0.30$ ($0.014$) & $0.25$ & $0$
    \end{tabular}}
    \caption{\modi{Parameters chosen for our methods on splits 1 and 2.}}
    \label{tab:split_params}
\end{table}

\begin{figure}[h]
    \centering
    \begin{tabular}{cc}
        \includegraphics[height=.2\textheight]{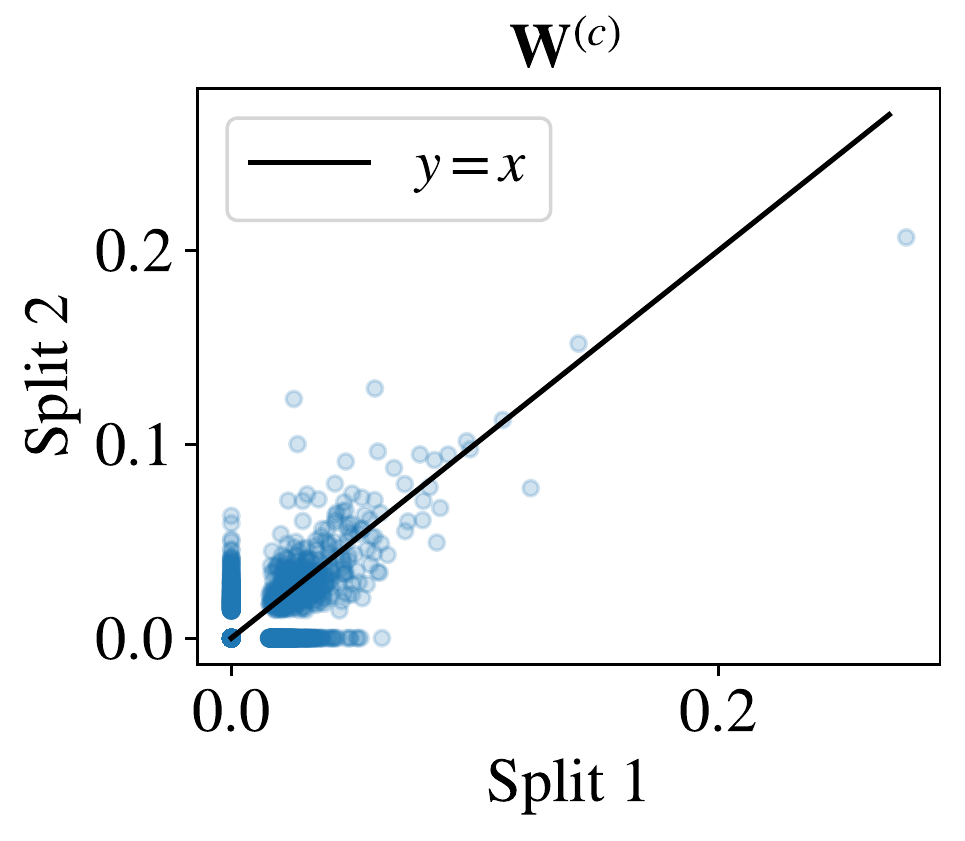} &
        \includegraphics[height=.2\textheight]{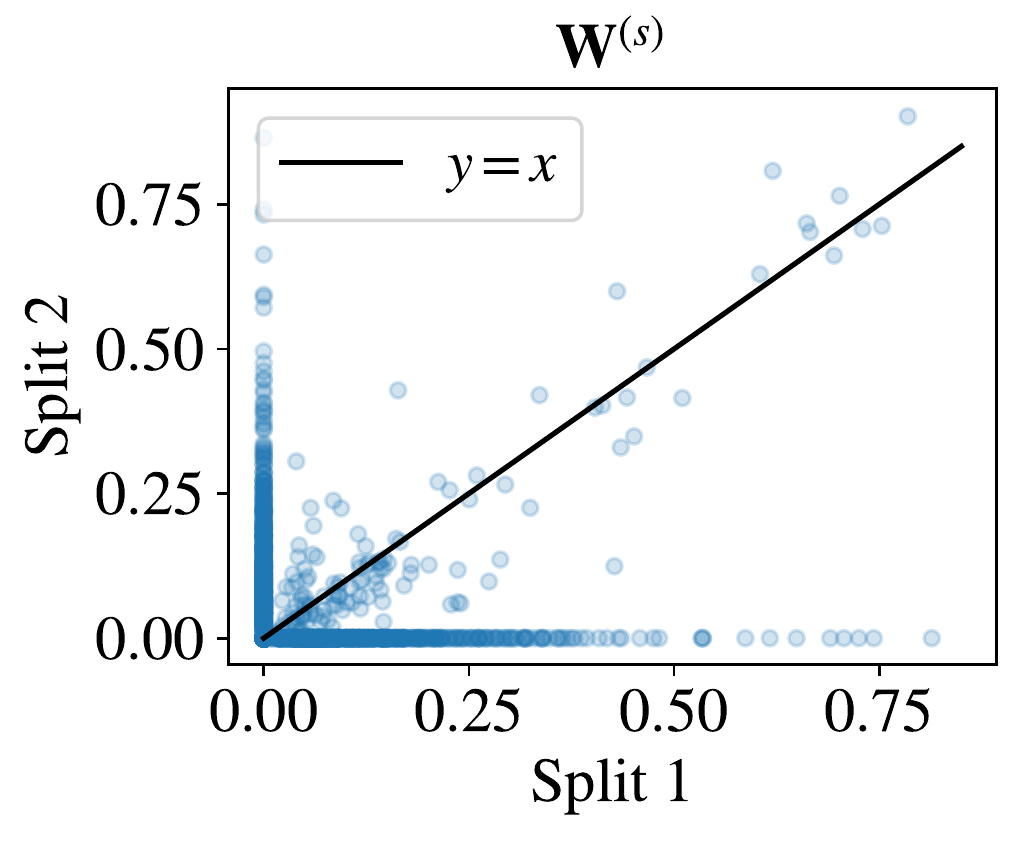}
    \end{tabular}
    \caption{Scatter plots of the complementarity scores (left) and the substitutability scores (right) from split 1 (x-axis) and 2 (y-axis), where the identity line $y=x$ is shown in black for reference.}
    \label{fig:2split_cs}
\end{figure}

In general, it is true in both complementarity and substitutability scores that if the values are larger than $0$ in both splits of the data, they are close to each other, see Fig. \hspace{-.4em}\ref{fig:2split_cs} where points $\{(x,y): x>0, y>0\}$ approach the identity line.
However, there are also many points have $x=0, y>0$, or $y=0, x>0$, i.e. the relationship is significant in one split but not the other. Since the same significance levels are chosen for both splits (see Table \ref{tab:split_params}), the insignificance from the other split stems from the change in customer  behaviour. Note that the substitutability scores are computed through the complementarity scores, thus there is also part of the points lying in either $x=0$ or $y=0$ obtained from sharing no complements\footnote{\footnotesize{There are $59.2\%$ of points $\{(x,y): x=0, y>0\}$ and $49.1\%$ of points $\{(x,y): x>0, y=0\}$ fall in this category.}}, which further caused by the changing behaviour of customers. 

\modi{Further to quantify how far the score values in one split from the other, we define the \textit{relative distance} $D$ between the scores $\{X^{(1)}_{ij}\}$ in split 1 and $\{X^{(2)}_{ij}\}$ in split 2 to be, 
\begin{align*}
    D = \frac{1}{n_p(n_p-1)}\sum_{i\ne j }D_{ij},\quad \text{where }
    D_{ij} = \frac{\abs{X^{(1)}_{ij} - X^{(2)}_{ij}}}{mean(X^{(1)}_{ij}, X^{(2)}_{ij})},
\end{align*}
and $n_p$ is the number of product nodes. Here, we use the arithmetic mean, $mean(x,y) = (x + y)/2$. The resulting relative distance of the complementarity and the substitutability scores are $0.13$ and $0.17$ respectively, thus both are relatively small.}

\modi{Finally, we compare the roles extracted from both splits. The complement roles from the two splits have NMI $0.74$ and AMI $0.62$, which indicates that the complement role structure from the two datasets largely agree with each other. While for the substitute roles obtained from the two sources, they have NMI $0.58$ but AMI $0.16$, where the difference between the two values is from the relatively large number of roles. Hence, there is slightly more change in the substitute role structure, which is expected since it also carries the changes from the complement relationship.} 

\modi{To conclude, in our chosen period, the temporal shifts will cause a certain amount of change in the score values, but will not cause dramatic change in the mesoscale role structure. Hence, as we mentioned in the very beginning, we find analysing the temporal change of our scores an interesting topic to explore in the future.}

\section{Formulation as unsupervised learning}
\label{sec:supp_unsupervised_learning}
The sales data can be seen as $n_t$ observations $(\mathbf{x}_1, \mathbf{x}_2, \dots, \mathbf{x}_{n_t})$ of a random $n_p$-dimensional vector $\mathbf{X} = (X_1, X_2, \dots, X_{n_p})$ having joint density $\mathbb{P}(\mathbf{X})$. The goal of unsupervised learning is to directly infer the properties of this probability density without the help of labels or degree-of-error for each observation \cite{hastie_unsupervised_2009}. Our goal here is specifically to find \modi{(i) (complements)} the item set $\mathcal{J}^{(c)}_i \subset \{1,2,..., n_p\}$ for each product $i$, s.t. $\forall j \in \mathcal{J}^{(c)}_i$, 
\begin{align}
	\mathbb{P}( X_j = 1 | X_i = 1),
	\label{equ:usl_com}
\end{align}
is significantly large; \modi{(ii) (substitutes)} the item set $\mathcal{J}^{(s)}_i \subset \{1,2,..., n_p\}$ for each product $i$, s.t. $\forall j \in \mathcal{J}^{(s)}_i,\ \exists j' \in \mathcal{J}^{(c)}_i$,  
\begin{align}
	\mathbb{P}( X_j = 1 | X_i = 0, X_{j'} = 1),
	\label{equ:usl_sub}
\end{align}
is significantly large. 

Instead of approximating the probabilities (\ref{equ:usl_com}) and (\ref{equ:usl_sub}) by the fractions of the corresponding observations in the data, which is not straightforward for the latter, we take a network perspective. In the product-purchase network, for the upcoming transaction node $l'$,
\begin{align*}
	\mathbb{P}( X_j = 1 | X_i = 1) = \mathbb{P}(a_{l'j} = 1 | a_{l'i} = 1) = \frac{\mathbb{P}(a_{l'j} = 1, a_{l'i} = 1)}{\mathbb{P}(a_{l'i} = 1)},
\end{align*} 
and similarly,
\begin{align*}
	\begin{split}
		\mathbb{P}( X_j = 1 | X_i = 0, X_{j'} = 1) &= \mathbb{P}( a_{l'j} = 1 | a_{l'i} = 0, a_{l'j'} = 1) \\
		&= \frac{\mathbb{P}( a_{l'j} = 1, a_{l'j'} = 1) - \mathbb{P}( a_{l'j} = 1, a_{l'i} = 1, a_{l'j'} = 1)}{\mathbb{P}( a_{l'i} = 0, a_{l'j'} = 1)}.
	\end{split}
\end{align*}

Then, to find the corresponding sets, we restrict set $\mathcal{J}^{(c)}_i$ to search values in
\begin{align*}
	\mathcal{I}^{(c)}_i = \{j: \mathbb{P}(a_{l'j} = 1, a_{l'i} = 1) > \alpha_m\},
\end{align*} 
i.e. product $i$ and $j$ are bought together significantly more frequently with significance level $\alpha_m$ (assumption 1 in Sect. \hspace{-.2em}\ref{sec:assumptions}); we also restrict set $\mathcal{J}^{(s)}_i$ to take values in 
\begin{align*}
	\mathcal{I}^{(s)}_i = \{j: \mathbb{P}(a_{l'j} = 1, a_{l'i} = 1) < \alpha_l,\ \mathbb{P}(a_{l'j} = 1, a_{l'j'} = 1) > \alpha_m,\ \exists j' \in \mathcal{I}^{(c)}_i \},
\end{align*}
i.e. product $i$ and $j$ are bought together significantly less frequently with significance level $\alpha_l$, but more frequently with some products in $\mathcal{I}^{(c)}_i$ (assumption 3 in Sect. \hspace{-.2em}\ref{sec:assumptions}). 

Next, we propose the measures to estimate the conditional probabilities. Based on random walks on networks, the estimated value of (\ref{equ:usl_com}) is taken to be the original directed measure, where $\forall j\in \mathcal{I}^{(c)}_i$,
\begin{align}
	\hat{\mathbb{P}}(a_{l'j} = 1 | a_{l'i} = 1) = sim_{od}(j,i)= \frac{\sum_{l=1}^{n_t}\frac{a_{lj}a_{li}}{d_l}}{\sum_{h=1}^{n_t}\frac{a_{hi}}{d_h}}.
	\label{equ:sim_od_usl}
\end{align}
We also modify the conditional probability to provide a symmetric version
\begin{align*}
	\mathbb{P}_m(j,i) = \frac{\mathbb{P}(a_{l'j} = 1, a_{l'i} = 1)}{\sqrt{\mathbb{P}(a_{l'i} = 1)\mathbb{P}(a_{l'j} = 1)}},	
\end{align*}
and approximate it by the original measure, where 
\begin{align}
	\hat{\mathbb{P}}_m(j,i) = sim_{o}(j,i)= \frac{\sum_{l=1}^{n_t}\frac{a_{lj}a_{li}}{d_l}}{\sqrt{(\sum_{h=1}^{n_t}\frac{a_{hi}}{d_h})(\sum_{h=1}^{n_t}\frac{a_{hj}}{d_h})}}.
	\label{equ:sim_o_usl}
\end{align}
The randomised versions are proposed in order to remove the specific noise effect from our estimates. 

The step to restrict the set of possible items is particularly important for estimating conditional probability (\ref{equ:usl_sub}), since we will remove the condition $X_i = 0$, and instead approximate $\mathbb{P}( a_{lj} = 1 | a_{lj'} = 1),\ \forall j'\in \mathcal{I}^{(c)}_i$, in an integrated way. The measure we propose directly for this purpose is the directed substitutability measure, where $\forall j\in \mathcal{I}^{(s)}_i$,
\begin{align}
    sim_{sd}(j,i) = \frac{\sum_{j'\in\mathcal{I}^{(c)}_j\cap \mathcal{I}^{(c)}_i}\modi{\min(\hat{\mathbb{P}}(a_{l'j} = 1 | a_{l'j'} = 1), \hat{\mathbb{P}}(a_{l'i} = 1 | a_{l'j'} = 1))}\hat{\mathbb{P}}(a_{l'i} = 1 | a_{l'j'} = 1)}{\sum_{p\in\mathcal{I}^{(c)}_i}\hat{\mathbb{P}}(a_{l'i} = 1 | a_{l'p} = 1)^2}.
    \label{equ:sim_sd_usl}
\end{align}
Note the $\hat{\mathbb{P}}(\cdot | \cdot)$ in Equation (\ref{equ:sim_sd_usl}) is an estimate of the probability in (\ref{equ:usl_com}), and we can consider, for example, estimates (\ref{equ:sim_od_usl}) and (\ref{equ:sim_o_usl}). We also modify the normalisation method to propose a symmetric version -- the substitutability measure, where
\begin{align}
	sim_{s}(j,i) = \frac{\sum_{j'\in\mathcal{I}^{(c)}_j\cap \mathcal{I}^{(c)}_i}\hat{\mathbb{P}}(a_{l'j} = 1 | a_{l'j'} = 1)\hat{\mathbb{P}}(a_{l'i} = 1 | a_{l'j'} = 1)}{\sqrt{(\sum_{p\in\mathcal{I}^{(c)}_i}\hat{\mathbb{P}}(a_{l'i} = 1 | a_{l'p} = 1)^2)(\sum_{p\in\mathcal{I}^{(c)}_j}\hat{\mathbb{P}}(a_{l'j} = 1 | a_{l'p} = 1)^2)}}.
	\label{equ:sim_s_usl}
\end{align} 

Now with the item sets and the corresponding estimates, we can output $\mathcal{J}^{(c)}_i$ and $\mathcal{J}^{(s)}_i$ for each product $i$, with any threshold/lower bound of the estimated probabilities. From our analysis of the sales data, $0.01$, $0.2$ are reasonable values of $\alpha_m$ and $\alpha_l$, respectively, and $0.01$, $0$ are sensible thresholds for the estimations (\ref{equ:sim_o_usl}) and (\ref{equ:sim_s_usl}), respectively. Following the path of network analysis, we can treat this process as a bipartite projection, and further analyse the community structure of the relationships determined by these probabilities, i.e. complements and substitutes, as in Sect. \hspace{-.4em}\ref{sec:results_salesdata}.
\end{document}